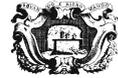

# Gamma-ray halos around pulsars: impact on pulsar wind physics and galactic cosmic ray transport

Elena Amato[1] · Sarah Recchia[2]

## Abstract
TeV haloes are a recently discovered class of very high energy gamma-ray emitters. These sources consist of extended regions of multi-TeV emission, originally observed around the two well-known and nearby pulsars, Geminga and PSR B0656+14 (Monogem), and possibly, with different degrees of confidence, around few more objects with similar age. Since their discovery, TeV haloes have raised much interest in a large part of the scientific community, for the implications their presence can have on a broad range of topics spanning from pulsar physics to cosmic ray physics and dark matter indirect searches. In this article, we review the reasons of interest for TeV haloes and the current status of observations. We discuss the proposed theoretical models and their implications, and conclude with an overlook on the prospects for better understanding this phenomenon.

## 1 Introduction

TeV haloes were serendipitously discovered in 2017 by the High-Altitude Water Cherenkov Observatory (HAWC). The HAWC collaboration reported the detection of Very High Energy (VHE, > 0.1 TeV) gamma-ray emission, at photon energy in the range 8–40 TeV, in coincidence with the nearby middle-aged pulsars Geminga and PSR B0656+14 [1]. Around each of the two pulsars, both located at a distance of ∼ 300 pc, the emission was reported to extend for 20–30 pc. Given the fact that pulsars are prominent factories of relativistic electron–positron pairs, the most straightforward interpretation of the emission is as a result of inverse Compton scattering (ICS) radiation produced by the pulsar-produced electrons and positrons.

✉ Elena Amato
elena.amato@inaf.it

✉ Sarah Recchia
sarah.recchia@inaf.it

[1] INAF-Osservatorio Astrofisico di Arcetri, Largo E. Fermi, 5, 50125 Firenze, Italy
[2] INAF-Osservatorio Astronomico di Brera, Via Bianchi 46, 23807 Merate, Italy

In the production of such multi-TeV photons, most of the electromagnetic spectrum is ruled out as target radiation for ICS due to the Klein–Nishina suppression of the cross-section, and the universal, and uniform, cosmic microwave background (CMB) is the only viable target, unless a local infrared radiation field with energy density larger than that of the CMB is present in the region (see, e.g., Ref. [2]). Indeed, this is the interpretation that accompanied the discovery article: upscattering of the CMB by the electrons and positrons. Within this interpretation, the nature of the emission implies a direct mapping of the spatial distribution of the emitting particles onto that of the radiation.

Moreover, given the $\gtrsim 100$ kyr age of these two sources, they likely lie outside the parent Supernova Remnant (SNR) and are currently releasing particles in the Interstellar Medium (ISM), while the 10 pc extension is likely much larger than the size of the Pulsar Wind Nebula (PWN). This is a crucial aspect that distinguishes "TeV haloes", the TeV emitters on which this review is focused, from the high-energy nebular emission or to the emission associated to pulsar that are still interacting with the SNR environment (see Sect. 4 for more details).

The HAWC Collaboration fitted the data with an electron distribution corresponding to loss-limited diffusive propagation from a central point like region of injection—the pulsar—finding that the data pointed to a diffusion coefficient 2–3 orders of magnitude lower than the average value in the Galaxy, when the latter is extrapolated from lower energy measurements of the ratio between secondary and primary CR nuclei (see, e.g., Refs. [3, 4], as reviews). This result immediately attracted much attention for at least three reasons: (1) its possible impact on the release of positrons by pulsars and their surrounding nebulae; (2) its possible impact on galactic transport of cosmic rays; (3) its possible impact on our understanding of diffuse $\gamma$-ray emission. We briefly introduce the three subjects here, deferring more details to later sections of this article. See also Ref. [5] for a recent review on the topic.

Let us start from the diffuse $\gamma$-ray emission, which is currently an extremely topical subject. Current measurements of the galactic diffuse emission at VHE and UHE (ultra high energies, $> 100$ TeV photon energies), show an excess [6] with respect to the standard expectation. The latter is based on the assumption that at these highest energies the diffuse emission is a result of CR nuclear collisions with interstellar gas, leading to neutral pion production and subsequent decay. This model, when supplied with the CR spectrum measured at the Earth and the distribution of target material inferred from measurements of atomic and molecular gas, accounts well for the level of emission detected in the outer parts of the Galaxy. However, the diffuse emission in the inner Galaxy is underpredicted.

In pulsar haloes, the emission is dominated by a local source over a size much larger than expected. So, if similar regions are widespread in the Galaxy, and in particular in the inner Galaxy, where sources are more numerous, the excess might be simply due to them, rather than indicating spatial inhomogeneity of the intensity of CR nuclei.

In addition, even if not too common, regions such as pulsar haloes need to be understood and correctly taken into account when estimating the background for VHE and UHE measurements. The estimate of the background level is preliminary to the identification of VHE gamma-ray sources, and large halos could hamper our ability at detecting new sources.

As far as the impact of TeV haloes on CR transport is concerned, if the interpretation of TeV haloes in terms of slow particle transport is correct, depending on their abundance and on the amount of suppression of the diffusion coefficient they entail, they might affect CR transport on galactic scales and impact the information we derive from it. In this respect, it is essential to understand how they arise and how frequently they develop.

More generally, it is worth noticing that the halo emission offers, at least at some locations, a window on particle propagation in an energy region of special interest, where the transport of CRs is not well constrained, either by other probes, nor in terms of theory. Cosmic ray transport through the Galaxy is ruled, presumably, by their resonant interaction with magnetic turbulence, present on top of a regular large-scale magnetic field.

The standard picture of magnetic turbulence in the Galaxy assumes that MHD turbulence is injected in the ISM primarily by Supernova (SN) explosions at an injection scale of tens of pc. The SN produced turbulence then cascades to smaller wavelengths, and in particular to the resonant scales (comparable to the particle Larmor radius, $r_L$) relevant for CR scattering. The effectiveness of scattering critically depends on the level and type of turbulent modes. In particular, if the cascade takes place according to an anisotropic distribution between wave modes that are parallel and perpendicular to the ambient magnetic field direction (for more details see Sect. 5 and references therein), very little power would be left in parallel modes, at the wavelengths that are relevant for scattering CRs with energies in the tens of TeV (0.01 pc). On the other hand, at these scales CRs are too few to effectively excite resonant waves themselves, contrary to what can happen in the 10 s–100 s GeV range. The combination of these two effects creates, in theory, a scattering gap, where it is not clear what rules CR transport. At the same time, while up to TeV energies, we gather information on CR transport from direct measurement of secondary to primary ratios, such diagnostic is not available beyond 1 TeV, a fact that makes TeV haloes especially precious to gain insight on this process.

Last but not least comes the impact of the existence of haloes on the release in the ISM of pulsar-produced positrons. A discovery that has raised much excitement in the last decade is the so-called "positron excess", one of the first results obtained when entering the era of high precision CR direct measurements with the spaceborne experiments PAMELA and AMS-02. In 2009, PAMELA [7] revealed that the ratio between cosmic positrons and electrons increases with energy above a few tens of GeV, in plain contrast with the predictions of models assuming that positrons are only produced as a result of CR interactions with the ISM during Galactic propagation. Indeed, the standard model of CR transport through the Galaxy assumes that relativistic particles propagate diffusively, by scattering with magnetic irregularities, and that the diffusion coefficient increase with particle energy as $D(E) \propto E^\delta$. In this picture, the diffusion coefficient is typically assumed as a global, spatially homogeneous parameter, and possible anisotropies, related to the presence of a large-scale Galactic field, are neglected. In the absence of losses, this results in a steady-state spectrum that for primary particles scales as $N_{\text{prim}}(E) \propto E^{-\gamma_{\text{inj}}-\delta}$, with $\gamma_{\text{inj}}$ the power-law index of the injected spectrum. For secondary particles, in the absence of losses and in the energy range in which the production cross-section can be approximated as energy independent, the injection index will coincide with the power-law index describing the steady-state spectrum of primaries, $\gamma_{\rm inj} + \delta$, leading to a steady-state spectrum: $N_{\rm sec}(E) \propto E^{-\gamma_{\rm inj}-2\delta}$. This framework is straightforward to expect $N_{\rm sec}/N_{\rm prim} \propto E^{-\delta}$. While both electrons and positrons undergo energy losses, due to synchrotron and inverse Compton emission above few tens of GeV, they are expected to be affected by losses in a similar way, so that the ratio between their fluxes should still decrease with increasing energy, $\Phi_{e^+}/\Phi_{e^-} \propto E^{-(\gamma_{\rm inj,p}-\gamma_{\rm inj,e^-}+\delta)}$, unless the injection spectrum of electrons is much softer than that of protons, with $\gamma_{\rm inj,e^-} > \gamma_{\rm inj,p} + \delta$.

The PAMELA finding of an increase with energy of $\Phi_{e^+}/\Phi_{e^-}$, soon confirmed by AMS-02 [8] with improved statistics and up to TeV energies, immediately prompted the particle and Dark Matter physics community for explanations based on the production of primary positrons by Dark Matter annihilation. At the same time, the CR Astrophysics community proposed explanations based on either non-standard propagation models [9–11] or on the relevance of well-known astrophysical sources of primary positrons, such as pulsars [12] and their nebulae [13].

Pulsars are fast rotating, highly magnetized neutron stars, that form as a result of the collapse of massive stars. As we discuss in the following, due to the large electric fields produced by the combination of high magnetization and fast rotation, pulsars are subject to intense electromagnetic torques that progressively slow down their rotation. The extreme electric field in the star vicinity creates ideal conditions for pair production, so that most of the rotational energy lost by the pulsar is converted into the production of a magnetized relativistic wind, whose material component is mostly made of electron–positron pairs.

As we discuss in more detail in the following, this outflow must slow down at some distance from the pulsar due to confinement by the surrounding medium, either the SN remnant or the ISM. Dissipation of the wind bulk energy happens at a termination shock and gives rise to the bright non-thermal sources known as PWNe [14]. The pulsar wind termination shock, in spite of occurring in a highly relativistic magnetized plasma, is a powerful accelerator: the PWN class prototype, the Crab Nebula, shows an efficiency of conversion of the wind bulk energy into accelerated particles $\gtrsim 20\%$ and maximum particle energy in the PeV range, the maximum that we believe achievable in Galactic sources [15].

The energetic electron–positron pairs that the pulsar produces are typically confined within the nebula before being released in the ISM. As long as the PWN material is confined by the SN remnant, the particles cannot leave the system, except possibly at the highest energies, where their Larmor radii become comparable with the system size. On the other hand, pulsars are a population of sources characterized by a high proper motion, with average speed in the 500 km/s range [16]. This implies that they will leave the SNR at some point during the Sedov–Taylor expansion phase of the latter. A rough estimate indicates that this will happen, for typical ISM conditions (particle number density in the ISM $n_{\rm ISM} \approx 1$ cm$^{-3}$), after few $\times 10^4$ yr. After this time, the pulsar wind confinement is only due to the supersonic pulsar proper motion through the ISM, which produces a bow shock in front of the pulsar in the direction of its motion [17]. At this stage, the accelerated particles are free to leave the system from the bow-shock tail and be released in the ISM. This is the picture that we believe to describe Geminga and Monogem.

Early calculations of the pulsar contribution to the positron excess [13] immediately showed that the latter could be easily accounted for by the release of pairs by systems in the bow-shock phase, such as Geminga and Monogem. When haloes were first discovered, they appeared as an important effect to take into account in these calculations, that were showing no need for positron sources other than pulsars otherwise. Of course, a reduced or reshaped pulsar contribution would reopen the quest for dark matter signatures in the $e^+ - e^-$ spectrum.

This article is organized as follows: in Sect. 2, we discuss in more detail the physics of pulsars and PWNe that is relevant for our purposes; in Sect. 3, we briefly review what we know about the population of accelerated particles in PWNe and their escape; in Sect. 4, we provide a review of the relevant observations; in Sect. 5, we discuss the connection between haloes and CR transport; in Sect. 6, we discuss the implications of haloes on the positron fraction; in Sect. 7, we provide a short summary and overview.

## 2 Physics of pulsar magnetosphere—pulsar wind

As we mentioned above, both Geminga and Monogem are relatively young pulsars. With typical surface magnetic fields $B_\star \sim 10^{12}$ G and rotation periods in the 10 s–100 s ms range, young pulsars work as powerful unipolar inductors, developing surface electric fields that far exceed the pull of gravity on charged particles at the neutron star surface. Electrons (and potentially also ions) are easily extracted from the star crust and distribute themselves so as to screen the local electric field in the star magnetosphere [18] (see Ref. [19] for a recent review). However, complete screening of the field everywhere around the star cannot be achieved with the charges extracted from the star alone. Huge potential gaps are bound to develop in the magnetosphere, where particles can be accelerated to very high Lorentz factors. The photons they emit while propagating along magnetic field lines are energetic enough to lead to pair creation in the intense magnetic and radiation fields present in the star vicinity. Copious pair production takes place, leading to increase the number of leptons initially extracted from the star by a factor $\kappa$, the so-called pulsar multiplicity, estimated in the range $\kappa \sim 10^3$–$10^5$ for young to intermediate pulsars [20].

The star magnetosphere is shaped by the fast stellar rotation and is divided in a corotating region and an open zone. The former is made by dipole magnetic field lines that extend to a maximum distance from the star less than the *light cylinder radius*, $R_L$, located at distance $c/\Omega_\star$ from the star rotational equator, with $c$ the speed of light ad $\Omega_\star$ the star rotation frequency. Particles in this region corotate with the star while sliding along magnetic field lines and never leave the system. At distances larger than $R_L$ corotation is no longer possible because it would require superluminal motion. The dipole field lines extending beyond $R_L$ become open and particles can leave the magnetosphere along these lines. The last closed field line defines the so-called polar cap, delimited by an angle $\theta_{\rm pc} \approx \sqrt{R_\star/R_L}$, with $R_\star$ the stellar radius. The potential difference between the magnetic pole and $\theta_{\rm pc}$ is the maximum available potential drop for the particles that leave the star. This can be worked out in perfect analogy with the physics of a Faraday disk and leads to a maximum achievable energy for accelerated particles [18]:

$$e\Phi_{\text{PSR}} = e\frac{B_\star \Omega_\star R_\star^2}{2c} \frac{R_\star}{R_L} = 6.6 \times 10^{12} \text{eV} \, B_{12} \, R_6^3 \, P_1^{-2} \tag{1}$$

where $e$ is the electron charge, $B_{12}$ is the neutron star surface magnetic field in units of $10^{12}$ G, $R_6$ the stellar radius in units of 10 km and $P_1$ the star rotation period in units of seconds.

The pairs that leave the star form a magnetized relativistic outflow that carries away most of the star rotational energy, while a subdominant part might be radiated in the form of gravitational waves. The star progressively spins down, releasing in the wind a luminosity $\dot{E} = I_\star \Omega_\star \dot{\Omega}_\star$, where $I_\star \sim 10^{45}$ g cm$^2$ is the star moment of inertia and $\dot{\Omega}_\star$ is the spin frequency time derivative. Current pulsar and pulsar wind models show that, as long as the dominant braking mechanism is of electromagnetic nature, independently of the details of how energy is carried away, $\dot{E}$ can be expressed in analogy with the luminosity of a radiating magnetic dipole [21]:

$$\dot{E} \approx \frac{B_\star^2 \Omega_\star^4 R_\star^6}{6c^3} = 10^{30} \text{erg/s} \, B_{12}^2 \, R_6^6 \, P_1^{-4} \, . \tag{2}$$

Equating the latter expression to the loss of rotational energy by the star, one easily finds that $\dot{E}$ evolves with time as $\dot{E} = \dot{E}_0/(1 + t/\tau_{SD})^2$ with the spin-down time, $\tau_{SD}$, defined as $\tau_{SD} = P_0 \dot{P}_0^{-1}/2$, with $P_0$ and $\dot{P}_0$ the pulsar initial period and its derivative, respectively. Since we do not have access to the initial values of $P$ and $\dot{P}$, it is customary to estimate $\tau_{SD}$ based on their current values: $\tau_{SD} \approx P \dot{P}^{-1}/2$. $\tau_{SD}$ is also used as an estimate of the pulsar age.

At large distance from the light cylinder, the pulsar outflow is well described as a cold MHD wind, expanding with constant speed. The wind is highly relativistic, with a bulk Lorentz factor estimated in the range $10^3 < \Gamma_w < 10^8$, and embeds a predominantly toroidal magnetic field. The energy partition in the wind must evolve with distance from the light cylinder: at $R_L$ the wind must be Poynting flux dominated, but a large fraction of its magnetic energy must be converted to particle kinetic energy before the termination shock (TS hereafter), where the wind is slowed down, its bulk energy dissipated and efficient particle acceleration occurs, as shown by the bright non-thermal emission starting immediately behind the shock [14] and references therein. While this phenomenology is best studied and understood in the case of the Crab Nebula, features similar to those observed in the class prototype are also seen in several other PWNe with spatially resolved X-ray emission [22, 23].

## 3 The particle population in PWNe

The PWN emission typically extends over a very broad range of photon energies, from the radio band to the VHE and UHE range. The emission mechanism is synchrotron radiation in the nebular magnetic field at energies up to the X-ray range, and ICS in the VHE range. The overall spectrum is well reproduced under the assumption that the accelerator produces a particle distribution in the form of a broken power-law: pairs with energy below $E_b \approx (0.5–1)$ TeV are characterized by an extremely hard energy distribution, $N(E) \propto E^{-\alpha_1}$ with $1 < \alpha_1 < 2$, while beyond $E_b$ the power-law index steepens to $\alpha_2 \approx 2.2–2.5$ [24, 25], more typical of shock acceleration in relativistic environments [26].

In fact, how particle acceleration occurs in PWNe is a big mystery. As mentioned above, the magnetospheric plasma reaches the TS as highly relativistic and magnetized, a condition that makes particle acceleration extremely challenging. Indeed, the most commonly invoked acceleration mechanism in Astrophysics, diffusive shock acceleration, is expected not to work in this context. The average magnetization, $\sigma_w$, namely the ratio between Poynting flux and particle kinetic energy flux, is expected to be not much lower than 1 at the pulsar wind TS (see, e.g., Refs. [27, 28] for reviews of the constraints on the subject derived from modeling of PWNe), while diffusive shock acceleration at relativistic shocks requires $\sigma_w \ll 1$ [26]. Even assuming that the plasma magnetization at the TS is highly inhomogeneous, with regions where $\sigma_w$ is sufficiently low as to allow diffusive shock acceleration, two of the observed properties of the particle population cannot be explained by this mechanism, namely the very hard spectral index below $E_b$ and the extreme maximum energies observed [29]. Alternative or additional proposals to explain the inferred particle spectrum, or at least some parts of it, are magnetic reconnection [30], turbulent acceleration [31], a mixture of the two in the vicinity of a shock [32, 33] and resonant cyclotron absorption in a ion-doped plasma [34]. What has become progressively clear is that likely a combination of different processes operating in different parts of the PWN need to be invoked [35].

Interestingly, whatever the acceleration process at work, there is an absolute limit to the particle energy that can be reached in PWNe, that is independent of any physics detail and only depends on pulsar parameters. Just as interesting is the fact that several PWNe seem to accelerate particles very close to this limit. The acceleration of leptons in relativistic sources is often limited by radiation losses. In general, $E_{\max} = \min[E_{\max,\text{abs}}, E_{\max,\text{loss}}]$, where $E_{\max,\text{abs}}$ refers to the maximum electric potential available in the accelerator, $E_{\max,\text{abs}} = e\Delta\Phi$, while $E_{\max,\text{loss}}$ is defined by the equality $t_{\text{acc}}(E_{\max}) = t_{\text{loss}}(E_{\max})$.

The acceleration potential $\Delta\Phi$ can be estimated as $\Delta\Phi = e\eta_E B L$, with $L$ the system size and $\eta_E$ the ratio between the electric and magnetic field in the source, $\eta_E = \mathcal{E}/B$. This implies

$$E_{\max,\text{abs}} = e\eta_E B L. \tag{3}$$

In highly conducting astrophysical plasma, the electric field $\mathcal{E}$ is typically $\mathcal{E} \approx (v_{\text{flow}}/c)B$, with $v_{\text{flow}}$ the characteristics speed of the plasma in the source, so that Eq. 3 is simply the Hillas criterion [36].

On the other hand, since magnetic fields in powerful accelerators usually entail an energy density that is larger than that of the radiation field, the dominant energy losses are typically due to synchrotron radiation, so that, assuming acceleration at the fastest possible rate ($t_{\text{acc}}(E) = E/(e\eta_E Bc)$), the loss-limited maximum energy reads

$$E_{\text{max,rad}} = m_e c^2 \left(\frac{6\pi e \eta_E}{\sigma_T B}\right)^{1/2} \approx 6 \text{ PeV } \eta_E^{1/2} B_{-4}^{-1/2} \quad (4)$$

where $m_e$ is the electron mass, $\sigma_T$ the Thomson cross section and $B_{-4}$ refers to the ambient magnetic field in the accelerator in units of $10^{-4}$ G.

In the case of PWNe, the wind is essentially non-radiative before the TS, as clearly shown by high resolution images that show a central dark region inner than the TS. While radiation losses are likely to be the limiting factor for particle acceleration in young powerful nebulae, most evolved nebulae, such as those powering haloes, will likely host magnetic fields that are well below $10^{-4}$ G at the TS (see, e.g., Ref. [37]).

The magnetic field at the TS can be related to the pulsar spin-down power writing the magnetic energy density as a fraction $\eta_B < 1$ of the wind ram pressure:

$$\frac{B_{TS}^2}{4\pi} = \eta_B \frac{\dot{E}}{4\pi c R_{TS}^2} . \quad (5)$$

While not straightforward to use for estimating $B$ unless the TS position is known, Eq. 5 can be used in Eq. 3 to obtain [38]:

$$E_{\text{max,TS}} = e \, \eta_E \, \eta_B^{1/2} \sqrt{\dot{E}/c} = 2 \text{ PeV } \eta_E \, \eta_B^{1/2} \dot{E}_{36}^{1/2} \quad (6)$$

with $\dot{E}_{36}$ the pulsar spin-down power in units of $10^{36}$ erg/s. The conclusion is that the maximum achievable energy at the TS only depends on the pulsar spin-down power $\dot{E}$. It is interesting to notice that $E_{\text{max,TS}}$ is also the maximum energy achievable in the pulsar magnetosphere, namely the potential drop at the pulsar polar caps, as one can easily check by comparing Eqs. 1 and 2.

Once we have established, through Eq. 6, the absolute maximum for the energy achievable in the pulsar-PWN system as a function of the pulsar $\dot{E}$, this immediately establishes that only pulsars with $\dot{E} > 10^{34}$ erg/s can produce the 1–100 TeV particles responsible for the formation of haloes such as those of Geminga's and Monogem's. At the same time, we do not expect haloes to be produced by too powerful pulsars, because these are likely young and still embedded in the SNR with which they were born.

As mentioned above, the pulsar population has an average velocity distribution around 500 km/s: during their life-time they will progressively move from their birthplace and evolve as illustrated in Fig. 1. From top left to bottom right, the figure illustrates the evolution of a pulsar through its SNR and out in the ISM. Initially the pulsar is near the center of the SNR and the latter is in free expansion. The PWN material is separated from the SN ejecta by a contact discontinuity and the reverse shock of the SN explosion is still propagating in the ejecta, outside of the contact discontinuity. As time advances, for typical parameters, the pulsar travels through its parent SNR while the latter enters the Sedov–Taylor phase. At this point, the reverse shock from the SN explosion propagates through the ejecta down to the PWN, compressing it.

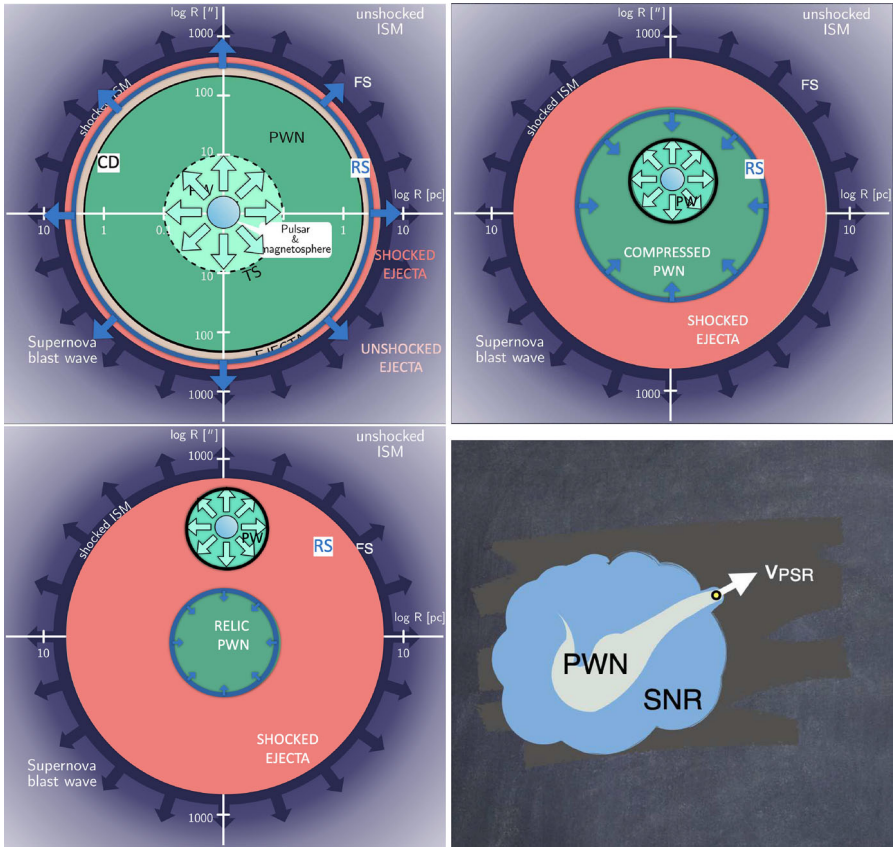

**Fig. 1** A cartoon illustrating the time evolution of a PWN. Top left panel: a young pulsar inflating a PWN while still close to its birth place; the PWN is embedded in a SNR which is still in the free expansion phase. Top right panel: the pulsar starts to be appreciably displaced from its birthplace and the PWN is compressed by the reverse shock propagating from the ejecta towards the center of the explosion; this is the reverberation phase, in which the PWN can undergo compression and re-expansion multiple times. Bottom left panel: the pulsar has moved away from the PWN during a compression phase and is now inflating a new PWN in the SN ejecta; a relic nebula has been left behind. Bottom left panel: the pulsar has left its parent SNR and its now moving through the ISM. The wind is confined by the ram pressure of the ISM, $\rho_{\rm ISM} v_{\rm psr}^2$

When the PWN pressure becomes so high as to contrast further compression, the PWN re-expands. The decreasing pulsar input (see Sect. 2) causes the cycle compression–re-expansion–compression to repeat few times [39]. During this phase it is possible that the pulsar motion takes the star out of the compressed PWN. The latter becomes causally disconnected from the parent pulsar, which will create a new nebula in its surrounding, leaving the old one as a relic [40]. The relic nebula will be left expanding in the SNR without any further energy input. It will become quickly depleted of high-energy particles, due to synchrotron and ICS losses, and will end up hosting only low-energy particles, emitting synchrotron radiation at radio wavelengths and producing gamma-rays in the TeV range through ICS. Relic PWN are likely the most numerous population of sources in TeV gamma-rays [37], and are relevant in this context because in some cases they might be confused with pulsar haloes. In this sense, it is important to remark the primary difference between these two source classes: while relic nebulae are within the SN ejecta, where the environment is likely highly turbulent, haloes develop in the ISM, where the turbulence level is expected to be close to the Galactic average. Finally, the pulsar will exit the SNR after a time $t_{\text{esc}}$ that can be calculated equating the pulsar displacement to the radius of the SNR during the Sedov–Taylor evolution of the latter [41]:

$$t_{\text{esc}} \approx 30 \, \text{kyr} \left( \frac{E_{\text{SN}}}{10^{51} \, \text{erg/s}} \right)^{1/3} \left( \frac{n_{\text{ISM}}}{1 \, \text{cm}^{-3}} \right)^{-1/3} \left( \frac{v_{\text{psr}}}{400 \, \text{km/s}} \right)^{-5/3} \quad (7)$$

where $E_{\text{SN}}$ is the SN explosion energy, $n_{\text{ISM}}$ the particle number density in the ISM and $v_{\text{psr}}$ the pulsar kick velocity. Once the pulsar has propagated out of the SNR and starts moving through the ISM, the confinement of its wind is due to the ram pressure of the latter, that in the pulsar reference frame reads $\rho_{\text{ISM}} v_{\text{psr}}^2$. If the pulsar motion is supersonic, a bow-shock PWN is then formed, with the PWN extending between the wind termination shock and the contact discontinuity with the shocked ISM. In this configuration the PWN electrons and positrons are not confined in the back of the nebula. Particles of all energies are free to escape the system from the tail. These are released in the ISM after suffering little or no losses depending on their energy [17]. The latter statement is based on the results from relativistic MHD simulations [42, 43] that show fast advection of most of the particles in the tail of the nebula. The detailed dynamics depend on several parameters, such as the pulsar wind anisotropy and magnetization and the inclination between the pulsar spin axis and the direction of the pulsar proper motion, but, in general, most of the plasma downstream of the termination shock flows away in direction opposite to the pulsar proper motion at a large fraction of the speed of light.

At the same time, both observations and numerical simulations show clear evidence that particles can also leave Bow Shock PWNe from regions other than the tail. On the observations side, we have evidence that the highest energy particles, close to the maximum available energy in the system, can leave Bow Shock PWNe from the head of the system, along narrow, X-ray bright trails [44–46]: this release is likely highly anisotropic [47], and seems to be accompanied by amplification of the interstellar magnetic field by a factor $\gtrsim 10$.

On the other hand, haloes themselves are evidence of nearly isotropic escape, involving again the highest energy particles. On the modeling side, studies have been performed tracking particle orbits on top of the electromagnetic field structure derived from 3D relativistic MHD simulations. These studies unveil some noticeable properties of the particle escape from Bow Shock PWNe. First, only particles within a factor of 20 of the maximum potential drop seem able to leave the system from the head of the bow shock in sizeable fractions: this is easily understood by noticing that, by definition, these particles have a Larmor radius that is comparable with the size of the termination shock, which is comparable, in turn, with the distance between the pulsar and the bow shock, the so-called stand-off distance, $d_0$, defined by the balance between the pulsar wind momentum flux and the ram pressure of the incoming ISM: $d_0 = \sqrt{\dot{E}/(4\pi c \rho_{\rm ISM} v_{\rm psr}^2)}$. Second, the distribution of escaping particles is progressively more isotropic with increasing energy. Only particles with energies very close to the pulsar potential drop ($E \in (0.1$–$1)\, E_{\rm max}$) form a quasi-isotropic outflow when escaping. In contrast, lower energy particles ($E \in (0.01$–$0.1)\, E_{\rm max}$) manage to escape only from selected regions where the PWN magnetic field reconnects with the interstellar magnetic field, and typically move along trajectories that are nearly aligned with the ISM magnetic field [48], namely, the particle pitch angles are distributed in a narrow range around 0. Finally, electrons and positrons in the lower energy range leave the system along different paths. When looking at particles in the range $E \in (0.01$–$0.1)\, E_{\rm max}$ the outflow is effectively charge separated: the escaping particles carry a net electric current. This is especially important in view of the processes that the propagation of these particles through the ISM can induce. The existence of a net current opens the possibility for the emergence of current driven instabilities, that can potentially lead to amplification of the ambient magnetic field by large factors [49].

In fact, the non-resonant streaming instability, vastly studied in connection with cosmic ray acceleration at Supernova Remnant shock, can in principle explain the large magnetic fields inferred from observations and modeling of X-ray trails [47, 50]. Of course, it is tempting to think that a similar mechanism might be responsible for the high turbulence level that halo observations seem to suggest.

We will discuss this hypothesis further in the following, after reviewing the relevant observations and the proposed models.

## 4 Halo observations

In the last 2 decades, a variety of experiments reported on a number of TeV $\gamma$-ray sources associated to pulsars or PWNe.

Most notably, Imaging Air Cherenkov Telescopes (IACTs), like H.E.S.S. [51], MAGIC and VERITAS, detected extended TeV emission around several young (spin-down age $\tau_{\rm SD} \lesssim 50$ kyr) and powerful (spin-down luminosity $\dot{E} \gtrsim 10^{35}$ erg/s) pulsars. In many cases, the spatial extension of the TeV emission is found to be substantially larger than what observed in other wavebands, and it often appears to be distinct from the X-ray counterpart of the associated PWN.

In many cases, the young pulsars associated with the detection are likely still confined in the parent SNR, of size $\approx 10$–$100$ pc depending on the age of the system and on the environment in which the SN exploded. Therefore, based on the evolution illustrated in Sect. 2, one may guess that the observed TeV emission results from a relic PWN, namely from the bubble of particles that detached from the pulsar during one of the compressions (by the SNR reverse shock) occurred during the reverberation phase. In this case, a synchrotron radio nebula should also be present in coincidence with the TeV emission, though possibly difficult to observe. Another possible interpretation, on the other hand, is that the emission is produced by particles that escaped the PWN (typical size $\approx 0.1$–$1$ pc) and are now interacting with the SNR environment.

A possible TeV halo of this kind has been reported by HAWC (HAWC J0635+070 [52]) in connection with the pulsar PSR J0633+0632 ($\tau_{SD} \sim 59$ kyr, $\dot{E} \sim 1.2 \times 10^{35}$ erg/s). These sources are surely very interesting from the point of view of investigating the evolution of PWNe, but, in both interpretations just proposed, the pulsar produced pairs probe the environment inside the remnant rather than the ISM, so that these sources do not help us gaining insight on CR propagation in the typical ISM. Moreover, electrons and positrons trapped in the SNR are likely to suffer significant energy losses and therefore unlikely to make a sizable contribution to the positron excess (see discussion in Sect. 6).

A somewhat different case is represented by the extended ($\approx 20$–30 pc) TeV haloes found around a few middle-aged ($\tau_{SD} \gtrsim 100$ kyr) pulsars, most notably around PSR J0633+1746 (Geminga) and PSR B0656+14 (Monogem) by HAWC [1], and, more recently, around PSR J0622+3749 by LHAASO [53]. Indeed, given their age and the typical pulsar speeds, $\sim 100$ km/s, these pulsars are likely outside the parent remnant and move supersonically in the ISM, creating a bow shock. As discussed in Sect. 2, particles produced by these sources can escape in the ISM, generate the observed TeV haloes, from which we can gain insight on the CR interstellar propagation at such high energies, and possibly make a dominant contribution to the CR positron excess detected at Earth.

Thus, in this review, we focus on this class of $\gamma$-ray haloes around middle-aged pulsars, and in this section, we summarize relevant information (see Table 1) related to these three sources and briefly discuss the prospects for future detections.

### Geminga–PSR J0633+1746

Geminga is a $\gamma$-ray and X-ray pulsar [54, 55] whose emission is detected from radio through $\gamma$-rays (see, e.g., Refs. [56, 57] and references therein). Located at a distance $d \approx 250$ pc from Earth [58, 59], it has characteristic spin-down age $\tau_{SD} \sim 340$ kyr, a period $P = 0.237$s and spin-down power $\dot{E} \sim 3.3 \times 10^{34}$ erg/s. Its proper motion was estimated to correspond to a transverse velocity of $\approx 200$ km/s, with a direction likely nearly transverse to the line of sight (LOS, hereafter) [58, 60]. Such speed implies that Geminga has moved $\approx 70$ pc away from its birthplace and has likely left the parent SNR.

*XMM-Newton* and *Chandra* observations revealed a peculiar "three-tail" PWN [57, 60, 61] with two nearly parallel X-ray tails of $\sim 2'$ extension ($\approx 0.2$ pc length) trailing behind the pulsar, and an axial tail of $\sim 0.05$ pc length resolved into a few individual segments [62]. The non-detection of H$_\alpha$ emission, together with the morphology of the tails, indicates that Geminga is moving through a rarefied (density $\lesssim 0.01$ cm$^{-3}$) and mostly ionized medium. Such conditions likely correspond to a hot ionized phase of the ISM, so that the Mach number of the bow shock may be moderate.

The interpretation of the X-ray tails, the outer ones especially, is not fully clear. While it is tempting to interpret the outer tails as tracing the contact discontinuity between the PWN material and the shocked ISM, their properties are inconsistent with the shape of the bow shock that one would expect for Geminga based on other observables: the bow-shock width is incompatible with the theoretical expectation, given

**Table 1** Properties of Geminga, Monogem and PSR J0622+3749 pulsars: period ($P$), spin-down luminosity ($\dot{E}$), characteristic age and distance ($d$, for PSR J0622+3749 the ∗ indicates that it is a pseudo-distance)

| | | $P$ [s] | $\dot{E}$ [$\times 10^{34}$ erg/s] | age [kyr] | $d$ [kpc] | $E_\gamma$ [TeV] | D (100 TeV) [$\times 10^{27}$ cm$^2$/s] | $\xi_{SD}$ (min) % |
|---|---|---|---|---|---|---|---|---|
| Geminga | HAWC | 0.237 | 3.26 | 342 | 0.25 | 8–40 | 3.2 | ∼ 5 |
| Monogem | HAWC | 0.385 | 3.8 | 11 | 0.288 | 8–40 | 15 | ∼ 1 |
| | | | | | | | 4.5 (combined fit) | |
| PSR J0622+3749 | LHAASO | 0.333 | 2.7 | 207.8 | 1.6* | > 25 | ≈ 7 | 40 $(d/1.6\,\text{kpc})^2$ |

For the HAWC and LHAASO analyses, we also report the photon energy range, the best-fit diffusion coefficient at $E = 100$ TeV (both the value for separate and combined Geminga+Monogem fits) and the minimal spin-down conversion efficiency, obtained with hard injection spectrum and cutoff in the range $E_c = 100$–150 TeV, as discussed in Sect. 5.1.1

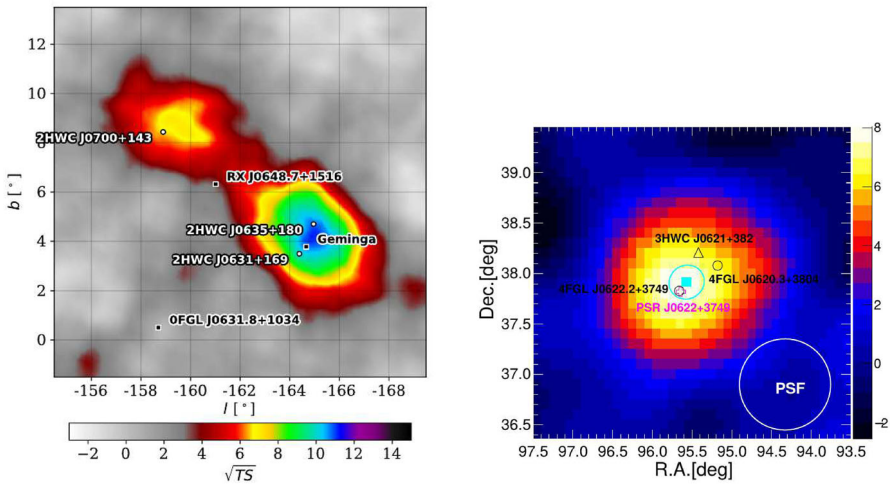

**Fig. 2** Left panel: HAWC significance map of Geminga and Monogem [1, 64] (© AAS. Reproduced with permission.) for $E_\gamma = 8$–40 TeV and for an extended source hypothesis, represented by a disk of radius of $2°$. Right panel: LHAASO significance map of PSR J0622+3749 [53] for $E_\gamma > 25$ TeV. The cyan square (circle) denotes the best-fit ($1\sigma$ range) of the location of the LHAASO source. The white circle shows the size of the LHAASO PSF (68% containment)

the pulsar speed and the low density, high temperature and highly ionized medium in which Geminga moves (see Ref. [62] and discussion therein). An alternative possibility is that the outer tails trace streams of pulsar wind particles escaping in the ISM, in analogy with the X-ray trails observed in association with the Guitar and Lighthouse nebula (see Sect. 3). In this case, they might be an integral part of the puzzle to solve to understand the extended TeV halo.

Geminga was first detected at TeV energies by Milagro [63] with a diameter of $\approx 2.8°$. Such detection was later confirmed by the HAWC collaboration [1], which reported the high significance detection of extended emission, on an angular scale of a few degrees, in the energy range $E_\gamma \in (8$–40) TeV. This range of photon energies corresponds to $\sim (10$–200) TeV leptons, if the emission is due to ICS on the CMB. Both the detected flux and estimated size of the emission were found to be comparable with those reported by Milagro. Such TeV source was named 2HWC J0635+180 and the corresponding significance map (taken from Ref. [64]) is shown in Fig. 2.

The HAWC collaboration modeled the emission, whose Surface Brightness (SB, hereafter) profile is reported in Fig. 3 (taken from Ref. [1]), under the assumption of isotropic diffusion of the electrons and positrons around the source.

Synchrotron and ICS losses were also included in the propagation model. The result was that, in order to explain the reported $\gamma$-ray SB, the diffusion coefficient around Geminga should be suppressed by 2–3 orders of magnitude compared to typical ISM values (see the discussion in Sect. 5.1). Their best-fit diffusion coefficient is $D(100\,\text{TeV}) \approx (3.2$–$4.5) \times 10^{27}\,\text{cm}^2/\text{s}$ (for individual fit and joint Geminga+Monogem fit) and is reported in Table 1 as a reference value for the following discussion. Recently, also H.E.S.S. confirmed the detection of an extended (at least 3°) emission around Geminga in the energy range $E_\gamma \in (0.5\text{--}40)$ TeV [65]. No significant indication was found of asymmetries or energy dependence in the $\gamma$-ray morphology. The diffusion coefficient at particle energy $\approx 100$ TeV used to fit the emission is close to that found by HAWC.

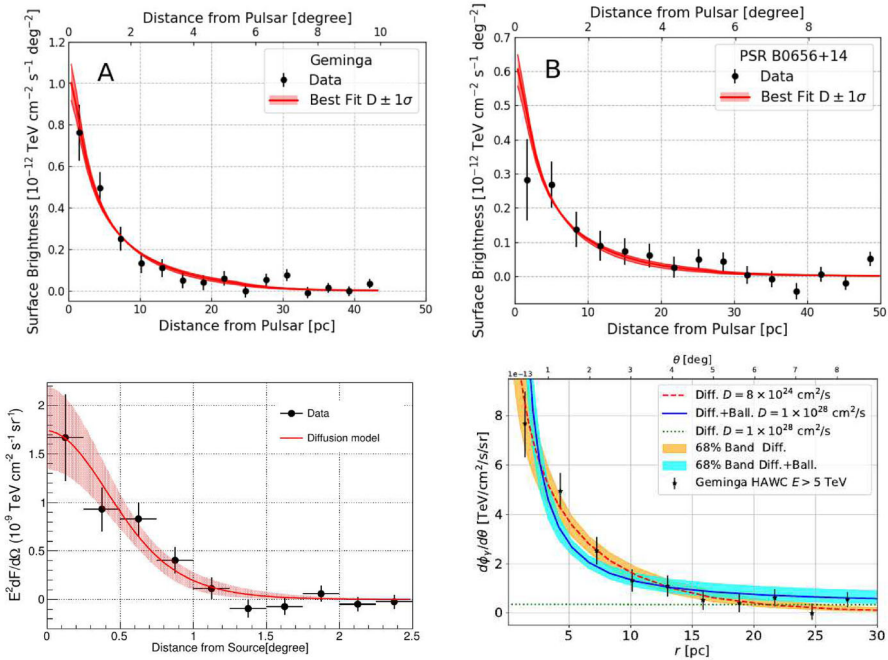

**Fig. 3** Top: $\gamma$-ray SB of Geminga (left panel) and of Monogem (right panel) by HAWC [1]. Solid red lines correspond to the best fit obtained by HAWC in the assumption of a 3D isotropic suppressed diffusion model. Bottom left panel: same as top panels for PSR J0622+3749 by LHAASO [53]. Bottom right panel: fit to the Geminga SB by [79] in the case of (i) 3D isotropic suppressed diffusion model (dashed red curve with orange band); (ii) 3D isotropic Galactic diffusion (dotted green line); (iii) 3D isotropic Galactic diffusion model with the inclusion of the ballistic propagation at early times (solid blue curve with cyan band)

Following these observations at multi-TeV energies, some authors [66, 67] investigated the 10 years *Fermi*-LAT data in the direction of Geminga and Monogem, in the framework of a spherical two-zone diffusion model (discussed in Sect. 6.1), with a suppressed diffusion coefficient up to some distance $\approx$ few $\times 10$ pc from the pulsar (in agreement with HAWC results) and the typical diffusion coefficient beyond that distance. While the theoretical considerations illustrated in Sect. 3 suggest that particles with energies in the multi-GeV range escape the nebula in a different way than particles with energies close to the pulsar potential drop, if isotropic diffusion outside the nebula is assumed, the formation of a GeV halo is not impossible, and the current resolution of gamma-ray instruments does not allow to discriminate the details of particle injection.

In the case of Geminga, the two analyses lead to opposite results, in that Ref. [66] finds no spatially extended emission, while Ref. [67] reports on a significant emission with a halo extended over the impressive size of $\approx 100$ pc. The implications of the combined HAWC and *Fermi*-LAT analyses on both the constraints on the particle injection spectrum and on the positron fraction were addressed in detail in the literature, and will be discussed in Sect. 6.

Finally, since the multi-TeV leptons responsible for the TeV $\gamma$-ray halos are expected to emit synchrotron X-ray photons ($h\nu_{\text{sync}} \approx 0.1$ keV ($B_{\text{ISM}}/\mu$G) $(E/10\,\text{TeV})^2$), Ref. [68] analyzed the *XMM-Newton* and *Chandra* data in a $600''$ region around Geminga to constrain the value of the magnetic field in the TeV halo. The authors analyzed, at the same time, the X-ray and the $\gamma$-ray HAWC data in the framework of the suppressed isotropic diffusion model used by HAWC, and obtained an upper limit for the diffuse X-ray flux at a level of $\lesssim 10^{-14}$ erg cm$^{-2}$ s$^{-1}$, corresponding to an upper limit $\lesssim 1\,\mu$G on the magnetic field. Such weak magnetic field (compared to typical $\gtrsim 1\mu$G values in the disk) could have relevant implications on the local CR transport conditions and/or on the local topology of field lines (see also Sect. 5).

### Monogem–PSR B0656+14

Monogem was discovered as a radio pulsar and has been extensively investigated through a variety of multi-wavelength observations (see Refs. [69, 70] and references therein). Located at a distance $d \approx 288$ pc from Earth [71], it has a characteristic spin-down age $\tau_{\text{SD}} \sim 110$ kyr, a period $P = 0.385$s and spin-down power $\dot{E} \sim 3.8 \times 10^{34}$ erg/s. Its proper motion corresponds to a transverse velocity of $\approx 60$ km/s and, based on the morphology of its putative PWN, is presumably directed close to the LOS [70], so that the actual speed may be of hundreds km/s. In addition, it is believed to be associated with the Monogem ring, a bright diffuse SNR with $\approx 12$–$13°$ radius [72].

Similar to the case of Geminga, in 2017 the HAWC collaboration reported on extended $\gamma$-ray emission in the range $E_\gamma = 8$–$40$ TeV around the Monogem pulsar. The source was named 2HWC J0700+143 and the corresponding significance map (taken from Ref. [64]) is shown in Fig. 2, while the SB (taken from Ref. [1]) is shown in Fig. 2. Also in this case, a suppressed diffusion coefficient in the source region was found, with $D(100\,\text{TeV}) \approx 4.5$–$15 \times 10^{27}$ cm$^2$/s (for joint Geminga+Monogem fit and individual fit, respectively) and is reported in Table 1.

In the case of PSR B0656+14, both analyses of *Fermi*-LAT data toward that region [66, 67] conclude against any significant emission.

### LHAASO J0621+3755–PSR J0622+3749

The LHAASO-KM2A experiment reported the detection of an extended source $\gamma$-ray emission at energies above 10 TeV: the source was named LHAASO J0621+3755 [53].

The source had not been detected before at VHE: neither the TeVCat [73] nor the second HAWC catalog [64] listed it. More recently the source 3HWC J0621+382 was reported in the third HAWC catalog [74], with an angular distance from the LHAASO source of $0.31° \pm 0.32°$.

In the GeV domain, in the vicinity of LHAASO J0621+3755, one can find the 4FGL J0622.2+3749 and 4FGL J0620.3+3804 *Fermi*-LAT sources [75]. The latter is classified as a blazar candidate in the 4FGL catalog, while the former is a $\gamma$-ray pulsar J0622+3749 discovered by *Fermi*-LAT [76].

With a period of $\sim 0.333$ s, a spin-down luminosity of $\sim 2.7 \times 10^{34}$ erg/s and a characteristic age of $\sim 207.8$ kyr, PSR J0622+3749 is a middle-aged pulsar similar to Geminga and Monogem. Currently, its distance is not well measured, and a pseudo-distance of $\approx 1.6$ kpc was inferred from the correlation between the $\gamma$-ray luminosity and the spin-down power for $\gamma$-ray pulsars [77].

Apart from the pulsed $\gamma$-ray emission, no counterparts of the pulsar in other wave bands have been found, in spite of looking for pulsed emission in radio, infrared, optical and X-rays [76]. Similarly, no clear evidence of extended emission was found at other wavelengths, including radio (see Ref. [53] and references therein).

Given the evidence for extended emission around Geminga in the GeV band found by Ref. [67], the authors of Ref. [53] also analyzed the *Fermi*-LAT data towards LHAASO J0621+3755 and did not find any significant extended emission. The reported upper limits were used to constrain [53] the injection spectrum and possible non-diffusive model of CR propagation [78].

## 5 Propagation

The ISM is embedded with magnetic fields, often composed of a large-scale coherent component, $B_0$, and a random component made of a variable mixture of magneto-hydrodynamic (MHD) fluctuations, both of the incompressible (Alfvénic) and compressible (magnetosonic) type (see Refs. [80, 81] for a review). Together with such modes, large-scale simulations have showed the appearance of conspicuous coherent structures and the development of intermittency (see, e.g., Refs. [82, 83]), which seems to have a relevant impact on the propagation of CRs [83, 84].

Turbulence can be injected on large scales ($L_{\text{inj}} \approx 1$–$100$ pc) by random plasma motions or by astrophysical sources, but also by CRs themselves through a variety of CR-driven instabilities. The turbulent component is characterized by the injection scale $L_{\text{inj}}$, the coherence length $L_{\text{coh}}$ (typically a fraction of $L_{inj}$), the root mean square field $B_{\text{rms}}$ and the relative normalization $\eta_B \equiv B_{\text{rms}}/B_0$. Moreover, it can be isotropic with respect to $B_0$, as in the case of Bohm, Kolmogorov and Kraichnan-type turbulence, or anisotropic, as in the case of Goldreich–Sridhar turbulence [85–90].

The CR transport in interstellar space is the result of the complex interactions of particles with such turbulent magnetic fields. The possible presence of a large-scale coherent field, differentiating the transport along and across the average large-scale field, makes the propagation anisotropic on scales smaller than $L_{\text{coh}}$ for any value of $\eta_B$ and even on larger scales for $\eta_B \ll 1$.

**Transport in the very close vicinity of the injection site** Here, we focus on the particle transport within a few $L_{\text{coh}}$ from the injection site. Once particles have been injected in a given region of space, their transport along the mean field is initially a ballistic gyromotion. Then, the interaction of particles with plasma waves at the resonant scale $k \sim 1/r_L$ (although non-resonant interaction may also play a role, Refs. [89–91]) leads to pitch-angle scattering and eventually to spatial diffusion along field lines.

When fluctuations are small compared to $B_0$, namely $\delta B/B_0 \ll 1$, the quasi-linear theory (QLT) of particle transport (see, e.g., Ref. [87] for a review) can be applied and the *parallel* diffusion coefficient (i.e., parallel to the large-scale magnetic field) can be expressed as

$$D_\parallel(E) = \left. \frac{4\pi c\, r_L(E)}{3 I(k_{\text{res}})} \right|_{k_{\text{res}}=1/r_L} \equiv \frac{1}{3}\lambda_{\text{mfp}}\, c, \qquad (8)$$

where $I(k_{\text{res}}) = \delta B(k_{\text{res}})^2/B_0^2$ is the normalized wave energy density calculated at the resonant wavenumber, and $\lambda_{\text{mfp}}$ is the particle mean free path. The scattering time $\tau_{\text{mfp}} \equiv \lambda_{\text{mfp}}/c$ is the timescale for the transition from a ballistic gyromotion to a diffusive motion.

$D_\parallel$ is generally energy dependent, as a result of the spectral properties of the turbulence [89] and of the possible presence of damping processes. In order to effectively scatter CRs, turbulence injected on large scales ($\gg r_L$) has to cascade to resonant scales ($\ll$ pc for sub-PeV particles) without being damped or becoming inefficient for scattering. For instance, the anisotropic cascade of Alfvénic turbulence makes it very inefficient at scattering CRs, while, if not damped, magnetosonic modes, whose cascade is isotropic, may dominate the CR diffusion [89, 90]. More recently, intermittency has also been shown to able to significantly contribute to CR scattering [83, 84].

On the other hand, CRs can effectively enhance their scattering along $B_0$, e.g. through the excitation of Alfvén waves at the resonant scale [92]. At the same time, plasma waves can also be damped though a variety of processes, such as the ubiquitous non-linear Landau damping, and ion-neutral damping, very effective when the medium is partially ionized (see, e.g., Ref. [93] and references therein).

The particle transport perpendicular to the mean field is less understood, but generally believed to be a combination of: (i) the random motion of field lines (FLRW), as due to turbulence on scales $\gg r_L$; (ii) the small-scale cross-field diffusion, whereby, due to scattering and drift, particles can jump from one field line to a nearby one [88, 94, 95]. Usually the latter process is characterized by a very small diffusion coefficient $D_\perp \ll D_\parallel$ (see, e.g., Ref. [85]), so that perpendicular transport is likely dominated by the FLRW. This process can be approximated, on scales comparable with $L_{\text{coh}}$ (see discussion by Refs. [85, 94, 96]), as a diffusion process characterized by a field line diffusion coefficient $D_m$ [95, 97],[1] with the value of $D_m$ depending on the type and level of turbulence, but typically of order of a fraction of $L_{\text{coh}}$ [85, 95, 97].

Due to the exponential divergence of neighboring field lines [94, 99], the combination of FLRW and small-scale cross-field diffusion tends to result in a perpendicular diffusive motion of particles. The diffusive behavior is finally established after particles have traveled a distance $\gtrsim L_{\text{coh}}$ in the turbulent field [85, 94, 96]. The resulting diffusion coefficient is typically $D_\perp \lesssim D_\parallel$, with the ratio tending to unity as the turbulence level increases.

---

[1] however, see also Ref. [98] for recent results on non-diffusive behaviors of field lines.

**Large-scale transport** In the discussion above we focused on scales below a few times the field $L_{\text{coh}}$. On such scales it is possible to identify a mean field and to effectively distinguish parallel and perpendicular transport even in the absence of a regular field component. Indeed, in the latter case, for particles whose Larmor radius is $\ll L_{\text{coh}}$, large-scale turbulent modes act locally as a mean field [85, 100]. Moreover, we mentioned that, in general, the CR diffusion can be anisotropic, with $D_\perp < D_\parallel$ when the level of turbulence is $\eta_B \lesssim 1$.

When the level of turbulence is increased, on sufficiently large scales ($\gg L_{\text{coh}}$) the propagation can be approximated as 3D isotropic diffusion (see, e.g., the discussion by Refs. [100, 101]) with

$$D_{\text{global}} \approx \frac{1}{3} L_{\text{coh}} v_{\text{eff}}, \tag{9}$$

where $v_{\text{eff}}$ is the effective velocity of particles in the direction of the local field (defined in each $L_{\text{coh}}$). Qualitatively, if particles undergo diffusive motion along the mean field, $v_{\text{eff}} \sim D_\parallel / L_{\text{coh}}$ and $D_{\text{global}} \sim D_\parallel / 3$. If particles move at speed $v$, one would get $v_{\text{eff}} = v$ and $D_{\text{global}} \sim L_c v$.

CR transport through the Galaxy is typically treated in the assumption of uniform and isotropic diffusion. The values of the CR interstellar diffusion coefficient, as inferred from CR data are (see, e.g., Ref. [102] for a review)

$$D(E) \approx D_0 E_{\text{GeV}}^\delta \, \text{cm}^2/\text{s}, \tag{10}$$

where $E_{\text{GeV}}$ is the particle energy in GeV, $D_0 \approx 10^{28} \text{cm}^2/\text{s}$ and $\delta \sim 0.3$–$0.6$, with $\delta = 1/3$ corresponding to a Kolmogorov-type turbulence and $\delta = 1/2$ to a Kraichnan-type turbulence. The resulting mean free path and scattering time are

$$\lambda_{\text{mfp}}(E_{\text{GeV}}) \approx 0.3 \, D_{0,28} E_{\text{GeV}}^\delta \, \text{pc} \tag{11}$$

$$\tau_{\text{mfp}}(E_{\text{GeV}}) \approx 1.0 \, D_{0,28} E_{\text{GeV}}^\delta \, \text{yr}. \tag{12}$$

Any model of TeV haloes needs to explain the currently detected sources, and in particular

1. the approximately spherical morphology of the emission.
2. the spatial extension of $\approx$ few $\times$ 10 pc.
3. the overall luminosity, which is bound to be a fraction $< 1$ of the pulsar spin-down.

In the rest of this section, we illustrate the main theoretical attempts that have been made in order to explain TeV haloes, discussing their advantages and drawbacks. Most of them focused on finding a plausible explanation for the inhibited diffusion found by HAWC (as anticipated in Sect. 4).

### 5.1 Isotropic diffusion

Given the approximately spherical geometry of the currently detected TeV haloes, most analyses, including that made by the HAWC and LHAASO collaborations [1, 53, 103, 104], interpreted the emission in the context of a 3D isotropic diffusion of CRs in the source region.

The resulting diffusion-loss equation for the particle distribution $f_e(t, r, E)$, as a function of the source age $t$, distance from the source $r$ and particle energy $E$, reads

$$\frac{\partial f_e}{\partial t} - \frac{D(E)}{r^2} \frac{\partial}{\partial r}\left(r^2 \frac{\partial f_e}{\partial r}\right) - \frac{\partial}{\partial E}(b(E) f_e) = Q_e(t, E)\delta(\mathbf{r}), \quad (13)$$

where $b(E) = -dE/dt$ is the energy loss rate and $Q_e(t, E)$ is the spectrum of leptons injected by the pulsar (or PWN) per unit time. For the latter, the normalization is typically set by assuming that a fraction $\xi_{SD}$ of the pulsar spin-down luminosity is channeled into CR $e^{\pm}$. As far as losses are concerned, for particle energy of $E \approx 10$–200 TeV, these are dominated by ICS on the CMB (scattering with all other relevant radiation fields is suppressed by Klein–Nishina effects) and by synchrotron on the background magnetic field, with a resulting loss time [105]:

$$\tau_{\text{loss}}(E) \approx 3 \times 10^4 \left(\frac{10\,\text{TeV}}{E}\right) \frac{1}{0.26 + 0.22\left(\frac{B}{3\,\text{uG}}\right)^2}\,\text{yr}. \quad (14)$$

Over such timescale, typically the pulsar spin-down luminosity does not change appreciably, so we can assume the injection as constant in time and obtain an approximate steady-state solution of Eq. 13 as:

$$f_e(t, r, E) \sim \frac{Q_0(E)}{4\pi D(E) r} \text{erfc}(r/r_d), \quad (15)$$

where $r_d(E)$ is the diffusion-loss length

$$r_d(E) \sim 2\sqrt{D(E)\tau_{\text{loss}}(E)} \approx 6 \sqrt{\frac{D}{10^{27}\,\text{cm}^2/\text{s}} \frac{\tau_{\text{loss}}}{10^4\,\text{yr}}}\,\text{pc}. \quad (16)$$

The CR distribution function exhibits a cutoff at $r \sim r_d$, due to the fact that particles lose all their energy within such distance. In order to reproduce the surface brightness of TeV halos, with an extension of $\approx 20$–30 pc, the diffusion coefficient should be $D(100\,\text{TeV}) \approx 10^{27}$–$10^{28}\,\text{cm}^2/\text{s}$, a value $\sim 100$–1000 times smaller than that obtained by extrapolating to high energies the Galactic diffusion coefficient deduced from CR observations (see Eq. 10). The assumption of such a small diffusion coefficient leads to radial profiles that are in very good agreement with both HAWC and LHAASO data, as shown in Fig. 3.

Interestingly, such values of the diffusion coefficient correspond to a diffusive mean free path $\lambda_{\text{mfp}} \lesssim$ pc, thus much smaller than the halo size. Moreover, as shown in a detailed study by Ref. [103], in the assumption of Kolmogorov or Kraichnan type of turbulence, such suppressed diffusion is compatible with a purely turbulent field with $L_{\text{coh}} \lesssim 1$–2 pc. More generally, as discussed above, the condition that $\lambda_{\text{mfp}}$ and $L_{\text{coh}}$ are much smaller than the halo size must be satisfied, if one is to describe the particle transport in the halo as due to 3D isotropic diffusion.

For Galactic values of $D(E)$, $\lambda_{\mathrm{mfp}}$ and $L_{\mathrm{coh}}$ become, respectively, comparable to and larger than the halo size, so that the picture of a pure 3D isotropic diffusion is no more adequate. If in Eq. 15 one naively increases $D$ to typical values, the result is, inevitably, a faint and spatially rather flat surface brightness profile, that does not agree with the data (see bottom right panel of Fig. 3).

In order to illustrate the impact of a large $\lambda_{\mathrm{mfp}}$, let us assume that particle propagation is spherically symmetric no matter the value of $D(E)$. Such (not very realistic) picture could apply either to the case of particles propagating in an unmagnetized medium and experiencing diffusion on some scattering centers whose average distance is $\lambda_{\mathrm{mfp}}$, or to the case in which field lines are radial in the source vicinity and particles diffuse along the lines. Neither of the two scenarios is very likely to occur in the ISM (see discussion below), but the assumption of radially symmetric transport allows us to investigate the limits of validity of Eq. 13 when $\lambda_{\mathrm{mfp}}$ is increased. As shown by Ref. [79], in this setup, the solution of Eq. 15 is substantially modified by freshly injected particles. Particles injected within a time $\approx \tau_{\mathrm{mfp}}$ have not reached the diffusive regime yet and contribute to the total particle distribution with a term $f_e \propto 1/r^2$. This term dominates the total distribution function up to distances from the source $r \sim \lambda_{\mathrm{mpf}}$. For Galactic values of the diffusion coefficient, for which $\lambda_{\mathrm{mpf}}$ becomes comparable to the size of the halo, the resulting SB is very close in shape to that of TeV halos (see Fig. 3), with the size of the emitting region being determined by the transition from ballistic to diffusive propagation rather than by losses. However, in order to reproduce the correct normalization of the SB, the required conversion efficiency of spin-down luminosity into accelerated leptons can substantially exceed 100%, since particles spread over a larger distance compared to the case of suppressed diffusion [79, 104].

On the other hand, also the hypothesis of spherical propagation may become inadequate for the typical Galactic diffusion of multi-TeV particles. Indeed, for typical turbulence models and ISM conditions, the galactic $D(E)$ corresponds to relatively large values of $L_{\mathrm{coh}}$ of tens pc (see, e.g., Ref. [86]). As discussed at the beginning of this section, and as shown by Ref. [103] by performing numerical simulations of particle transport in random fields, when $L_{\mathrm{coh}}$ is increased, particles injected by a source of size $R$ will trace a magnetic flux tube of the size of the source up to distances $\lesssim L_{\mathrm{coh}}$. Therefore, their distribution remains rather filamentary up to such distance, resulting in an increasingly asymmetric SB with increasing $L_{\mathrm{coh}}$. This can be seen in Fig. 4 (taken from Ref. [103]), where the $\gamma$-ray SB around Geminga was computed for values of $L_{\mathrm{coh}} = 0.25, 5, 10, 40$ pc. While for $L_{\mathrm{coh}} \lesssim 5$ pc the SB is well spherical, it becomes more and more asymmetric and filamentary for larger values. This result was used by the authors to constrain the value of $L_{\mathrm{coh}}$ in the surroundings of Geminga: their finding was $L_{\mathrm{coh}} \lesssim 5$ pc with a preferred value $L_{\mathrm{coh}} \sim 1$ pc. Such values are not incompatible with those found in the ISM $L_{\mathrm{coh}} < (1\text{–}100)$ pc, although at the low end of the range.

In summary, analyzing TeV haloes in the framework of pure 3D isotropic diffusion contains the implicit assumption that the field around the source is sufficiently turbulent, so that $L_{\mathrm{coh}}$ is substantially smaller than the size of the emitting region. In typical models of isotropic turbulence, such as in the Kolmogorov and Kraichnan cascades, this also implies a suppressed diffusion coefficient compared to typical Galactic values. The case of a field with larger coherence length, and of a larger diffusion coefficient,

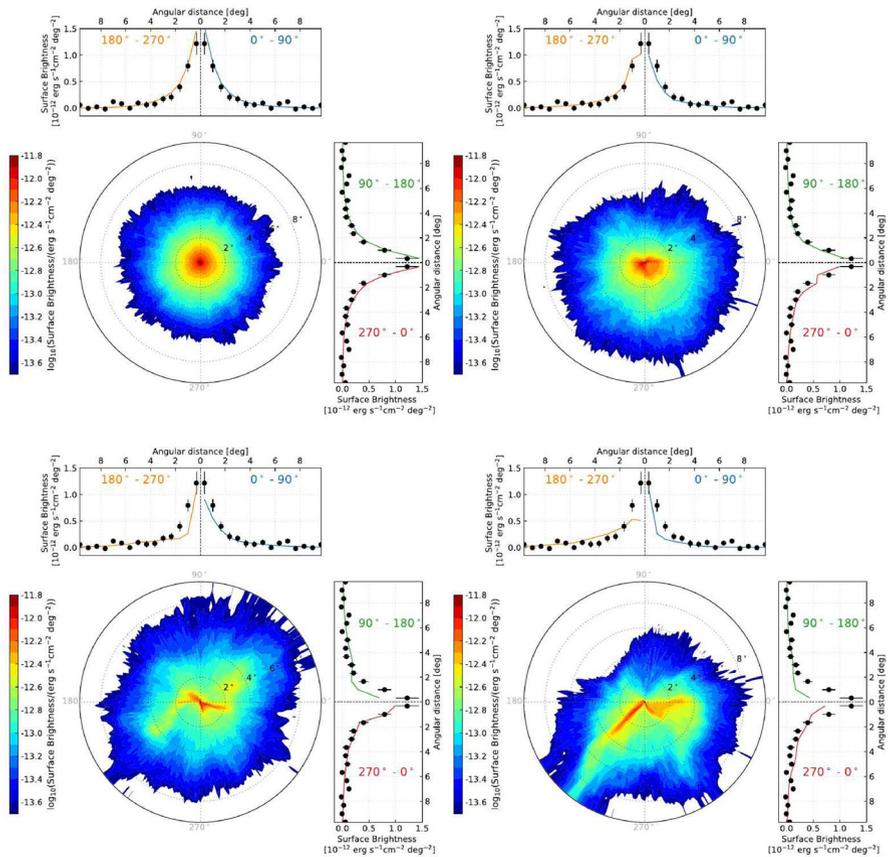

**Fig. 4** Taken from Ref. [103], the figure shows the $\gamma$-ray SB in polar coordinates for different values of the field coherence length: $L_{\rm coh} = 0.25$ pc (top left panel), $L_{\rm coh} = 5$ pc (*top right panel*), $L_{\rm coh} = 10$ pc (*bottom left panel*), $L_{\rm coh} = 40$ pc (bottom right panel). The corresponding SB profile, integrated over a quadrant and compared with the HAWC data, is shown in every panel at the top and right-hand side of every plot

cannot be correctly addressed in this setup, and requires a dedicated investigation, as shown later on. The scenario of small $L_{\rm coh}$ and suppressed diffusion naturally explains the rather spherical morphology and the size of TeV haloes (in terms of the diffusion-loss length), with a luminosity in particles corresponding, typically, to a fraction of the pulsar spin-down power $\lesssim 10s\%$ (this fraction may vary depending on the assumed particle injection spectrum).

The open question is the possible physical origin of such a scenario. A possibility suggested in the literature is that some pulsars can be located in regions with a much higher turbulence level than the average ISM. For instance, Ref. [106] proposed that TeV haloes could develop in the post-shock region of the (old) parent SNRs. This idea was proven to be viable in the case of Geminga provided that $\sim 10\%$ of the SNR energy is transferred to turbulence injected on a scale $\sim 10$ pc. In that work, however, adiabatic losses and the effects of possible damping were not taken into account. In addition, the pulsar is assumed to be still inside the parent remnant even at ages $\gtrsim 100$ kyr, which implies tight constraints on the SN environment and pulsar speed.

Other studies suggested that multi-TeV protons that escaped from SNRs may suppress the diffusion coefficient within tens pc from the remnant and even excavate bubbles in the surrounding medium [107]. Alternatively, pulsars located inside superbubbles may experience a level of turbulence substantially larger than in the unperturbed ISM (see, e.g., Ref. [108] and references therein). For all these scenarios, it is not fully clear whether the level of generated turbulence, its survival time and spatial extension are adequate to explain the detected TeV halos.

In any case, one may wonder what is the filling factor of such turbulent regions, the probability that a pulsar ends up within one of those, and whether they could affect the overall CR propagation and the production of secondaries [109].

### 5.1.1 Injection spectrum and acceleration efficiency

For the TeV haloes of Geminga, Monogem and PSR J0622+3749, the combination of their multi-TeV $\gamma$-ray flux and the upper limits (or possible detection) found in the multi-GeV domain (see Sect. 4), provide constraints on the conversion efficiency of the pulsar spin-down luminosity into accelerated leptons, and on the shape of the injection spectrum (see also Sect. 6.1).

Focusing on the 3D isotropic (suppressed) diffusion scenario, analyses of the HAWC and LHAASO data adopted for the pairs either a power-law (plus cutoff) injection spectrum:

$$Q_e \propto \left(\frac{E}{\text{GeV}}\right)^{-\alpha} e^{-E/E_c}, \qquad (17)$$

or a broken power-law, with a hard spectrum ($1 < \alpha_1 < 2$) below the break $E_b \approx 0.5$–1 TeV and a steep spectrum ($\alpha_2 \approx 2.2$–2.4) above the break, in agreement with observations of PWNe (see Sect. 2).

Spectral shapes that maximize the energy input in the multi-TeV leptons (e.g. a single hard power-law with $\alpha \lesssim 2$ and $E_c \sim 150$ TeV) provide a lower limit for the acceleration efficiency required to power the detected TeV haloes and reduce the $\gamma$-ray emissivity in the *Fermi*-LAT band.

For instance, the HAWC collaboration [1] adopted a single power-law spectrum for Geminga and Monogem, with particles injected in the range 1 GeV $-500\,TeV$. For the former, they found a slope $\alpha_{\text{Geminga}} \sim 2.34$ and an efficiency $\xi_{\text{Geminga}} \sim 40\%$, while for the latter $\alpha_{\text{Monogem}} \sim 2.14$ and $\xi_{\text{Monogem}} \sim 4\%$. Notice that with spectra steeper than $\alpha = 2$, most of the energy is in the low-energy part of the spectrum (in this case at 1 GeV), rather than in the multi-TeV range. The impact of choosing a different spectrum can be appreciated by comparing the Geminga analysis by HAWC with that by Ref. [104], which used a similar diffusion coefficient but assumed a very hard spectrum, with $\alpha_{\text{Geminga}} = 1$ and $E_c = 133$ TeV, and found $\xi_{\text{Geminga}} \sim 5\%$.

As for PSR J0622+3749, the LHAASO collaboration found $\xi_{\text{J0622}} \sim 40\%(d/1.6\,\text{kpc})^2$ with $\alpha = 1.5$ and $E_c = 150$ TeV. This is a rather large efficiency, compared to the case of Geminga and Monogem, considering the hard injection spectrum assumed. However, it is important to keep in mind that in the case of PSR J0622+3749 the distance is not much constrained and that $d \sim 1.6$ kpc is a pseudo-distance [77].

For the sake of future discussion, in Table 1, we report, for each of the three sources, the typical values found for the acceleration efficiency, under the assumption of 3D suppressed diffusion and hard injection spectrum with cutoff $\approx$ 100–150 TeV. Such values should be considered as a sort of lower limit for the efficiency.

### 5.2 Anisotropic diffusion

In an attempt to reconcile the small extension of TeV halos with the typical Galactic diffusion coefficient, Refs. [110, 111] proposed a model of anisotropic diffusion. Indeed, as discussed at the beginning of this section, when the level of turbulence is $\eta_B < 1$, the diffusion coefficient perpendicular to the mean field can be substantially smaller than that along $B_0$. In such scenario, $D_\parallel(E)$ can be close to typical Galactic values while $D_\perp$ could be suppressed. If the mean field around a source is roughly aligned with the LOS, the resulting TeV halo would appear compact.

For such scenario to work, two conditions need to be met: (i) in order for $D_\parallel$ to match the average Galactic value, $L_{coh}$ should be of the order of 10 s pc [86, 89] and $\eta_B \lesssim 0.5$; (ii) in order for the TeV halo to be approximately spherical, the inclination angle between the mean field and the LOS should be $\phi \lesssim \tan^{-1}\sqrt{D_\perp/D_\parallel}$. If the suppression needs to be by a factor 100–1000, then $D_\perp \lesssim D_\parallel/100$ and $\phi \lesssim 5°$ (see, e.g., Ref. [112]).

These qualitative arguments were confirmed by Refs. [110, 111], who studied the anisotropic propagation of leptons around Geminga in the assumption that

$$D_\perp = M_A^4 D_\parallel, \tag{18}$$

as found in the works by Ref. [101], where $M_A = \delta B/B_0$ is the Alfvénic Mach number of the turbulence at the injection scale ($M_A = \eta_B$ as defined at the beginning of Sect. 5). In the top panel of Fig. 5, we report the results by Ref. [110]. These authors computed the expected SB of Geminga for the cases $M_A = 0.1, 0.2, 0.3$ and inclination angle $\phi = 0°, 5°$ between the mean field and the LOS, with $D_\parallel(E) = 3.8 \times 10^{28}(E/\text{GeV})^{1/3}$ cm$^2$/s, which corresponds to $D_\parallel(100\,\text{TeV}) = 1.2 \times 10^{30}$ cm$^2$/s. In comparison, the value reported for Geminga by the HAWC analysis is $D(100\,\text{TeV}) \approx 5 \times 10^{27}$ cm$^2$/s.

For $\phi = 0°$, the SB is obviously symmetric around the position of the pulsar (assumed to be located at the origin of the coordinate system). The value of $M_A$ controls the compactness of the TeV halo, which, given the choice for $D_\parallel$, for $M_A = 0.1$ is too small ($D_\perp/D_\parallel = 10^{-4}$ and $D_\perp(100\,\text{TeV}) \sim 1.2 \times 10^{26}$ cm$^2$/s ), while for $M_A = 0.3$ ($D_\perp/D_\parallel \sim 0.008$ and $D_\perp(100\,\text{TeV}) \sim 10^{28}$ cm$^2$/s) is too large. Instead, for $\phi = 5°$ and $M_A = 0.1$ the SB is asymmetric, since the LOS on one side of the pulsar encounters more leptons than the LOS on the other side. The asymmetry becomes progressively less pronounced with increasing $M_A$. This can be seen also in the bottom panel of Fig. 5, which shows the gamma-ray emissivity map for photons $E_\gamma = 20$ TeV computed by Ref. [111] for $\phi = 90°$ and $M_A = 0.2, 0.5, 0.8$.

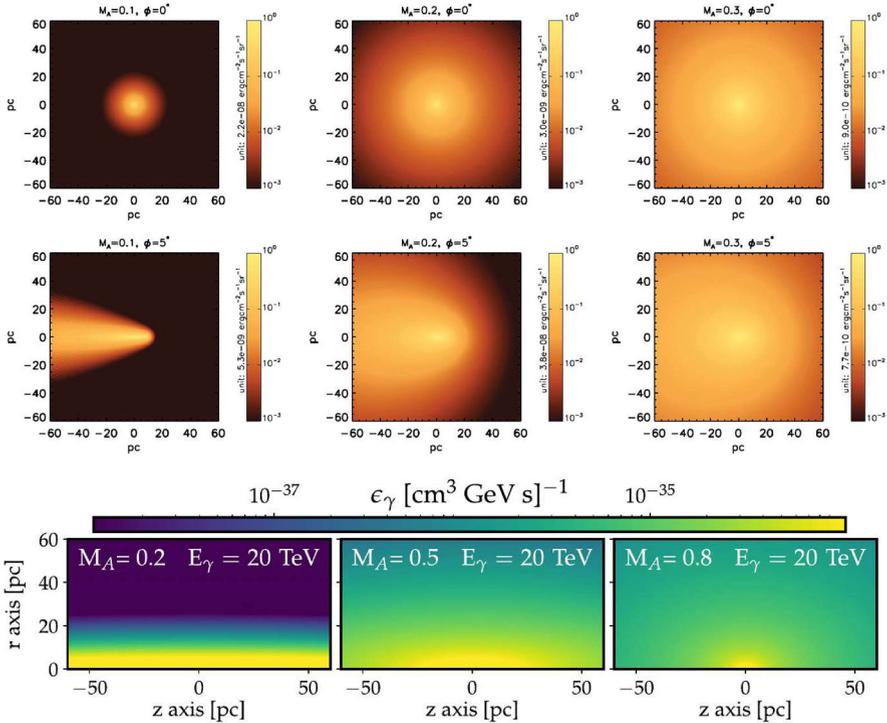

**Fig. 5** Top panel: $\gamma$-ray SB in the range $E_\gamma = 8$–40 TeV for different values of the Alfvénic Mach number, $M_A$, and of the angle $\phi$ between the mean field and the LOS (taken from Ref. [110]). Center panel: $\gamma$-ray emissivity map at $E_\gamma = 20$ TeV for different values of $M_A$ and fixed angle $\phi = 90°$ (taken from Ref. [111])

Based on these analyses, it was found that the conditions on the inclination angle and $D_\perp/D_\parallel$ to reproduce the HAWC data are rather stringent. This would not be a problem for a single source. For instance, Ref. [110] argued that a small inclination angle may also reconcile the rather low value of the magnetic field in the region of Geminga ($< 0.8 \mu$G), as inferred from X-ray observations [68], with typical Galactic values (say $\approx 3 \mu$G).

On the other hand, this interpretation becomes less likely as the number of detected spherical haloes increases, since the invoked scenario would rather predict elongated structures of TeV emission around middle-aged pulsars. One could be tempted to argue that even only a handful of haloes, if 100% spherical, is sufficient to rule the anisotropic diffusion hypothesis out. However, there could be observational biases (related to how the data analysis is performed) favoring the detection of spherical structures, and in addition Geminga and Monogem are not far from one another, so that in principle the two haloes could be developing within the fame magnetic flux tube. For instance,

Ref. [113] discussed the prospects for future detection of TeV haloes with LHAASO in the context of anisotropic CR transport.

As a final remark, a caveat of the illustrated models is the assumption that in the perpendicular direction propagation becomes diffusive immediately after injection from the source. Instead, as discussed above, and confirmed by simulations, the regime of perpendicular diffusion applies when particles have traveled a distance along the mean field $\approx$ a few $L_{\rm coh}$: this sets a spatial and temporal scale for reaching the perpendicular diffusive regime. For $L_{\rm coh}$ comparable to the size of the halo, this aspect has to be taken into account and may affect the conclusions reported above.

### 5.3 Cosmic ray self-generated diffusion

Gradients in the CR density, as those found in the proximity of SNRs or between the Galactic disk and halo, produce the excitation of Alfvén waves at the scale of the Larmor radius of particles that move along the CR gradient. This self-generated turbulence can dominate the particle transport in the Galaxy up to hundreds of GeV [114, 115] and may produce regions of suppressed diffusion around SNRs [93, 107, 116–119].

Based on these results, it was proposed that the same mechanism could produce a suppression of the diffusion coefficient around middle-aged pulsars [120, 121]. Following similar models to those proposed in the case of SNRs, these works considered the time and space-dependent evolution of the distribution function of the pulsar injected $e^{\pm}$ coupled to the evolution of the Alfvén waves assumed to determine the scattering coefficient. The equation for the wave evolution included wave generation by CR streaming instability and non-linear wave damping. The streaming instability was assumed to be seeded by the $e^{\pm}$ pairs themselves, injected in the ISM by a Geminga like pulsar.

A key feature of such models is that, up to a coherence length from the source, assumed to be $L_{\rm coh} \sim 100$ pc, the CR and wave propagation is considered to take place in a flux tube of transverse size $R \sim 1$ pc, roughly corresponding to the size of the PWN. Beyond such distance the propagation is assumed to quickly become 3D (spherical) and as soon as this happens, the CR gradient quickly decreases and the excitation of waves becomes negligible.

Therefore, the propagation within a distance $L_{\rm coh} \sim 100$ pc from the source can be treated in a 1D diffusion setup. In such scenario, the transport equations for CRs and waves depend on the distance along the flux tube, $z$, and read

$$\frac{\partial f_e}{\partial t} + v_A \frac{\partial f_e}{\partial z} - \frac{\partial}{\partial z}\left[D(p,z,t)\frac{\partial f_e}{\partial z}\right] - \frac{dv_A}{dz}\frac{p}{3}\frac{\partial f_e}{\partial p} + \frac{1}{p^2}\frac{\partial}{\partial p}\left[p^2 \frac{dp}{dt} f_e\right]$$
$$= Q_e(p,z,t) \qquad (19)$$
$$\frac{\partial I}{\partial t} + v_A \frac{\partial I}{\partial z} = (\Gamma_{\rm CR} - \Gamma_D)\, I(k,z,t). \qquad (20)$$

Here, $p$ is the particle momentum, $k$ is the wavenumber and $v_A = B_0/\sqrt{4\pi m_i n_i}$ is the Alfvén velocity defined on the background (unperturbed) field, $B_0$, and adopting ion mas density $m_i n_i$. $I(k, z, t) \equiv (\delta B(k)/B_0)^2$ corresponds to the wave energy density at wavenumber $k$.

The different terms in the transport equation are: on the left hand side, time derivative of the distribution function, advection with the waves, diffusion, adiabatic and energy losses due to synchrotron and ICS emission; on the right-hand side, injection by the pulsar (or PWN).

This system of equations is complemented by the self-generated diffusion coefficient, which can be computed as in Eq. 8 with $I(k_{\rm res})$ being the wave energy density at the resonant wavenumber $k_{\rm res} = 1/r_L$, and by the growth rate of waves due to the CR streaming instability, given by

$$\Gamma_{\rm CR}(k) = \frac{16\pi^2}{3} \frac{c v_A}{I(k) B_0^2} \left[ p^4 \frac{\partial f_e}{\partial z} \right]_{p_{\rm res}}, \tag{21}$$

where the term in square brackets is the gradient of CRs in resonance with waves of wavenumber $k$.

As for damping, both studies included the cascade of waves to smaller scales due to non-linear wave-wave coupling:

$$\Gamma_{\rm NLD}(k) = c_k v_A \begin{cases} k^{1/2} I^{1/2} & \text{(Kolmogorov)} \\ I & \text{(Kraichnan)} \end{cases} \tag{22}$$

in the Kolmogorov and Kraichnan phenomenology (with $c_k \simeq 0.052$). Other potentially relevant damping mechanisms are ion-neutral damping, non-linear Landau damping and turbulent damping (see, e.g. Ref. [93] and references therein) and were briefly discussed by Ref. [122].

The results of this model in terms of the time- and space-dependent suppression of the diffusion coefficient at particle energies of $\sim 10$–$200$ TeV critically depend on:

1. the transverse size $R$ of the flux tube, which controls the CR gradient with a dilution factor $R^2$
2. the conversion efficiency $\xi_{SD}$ of the spin-down luminosity into leptons, the spectral index $\alpha$ and cutoff energy $E_c \approx 100$ TeV of the injection spectrum ($Q_e \propto E^{-\alpha} e^{-E/E_c}$), which together fix the amount of energy channeled into multi-TeV $e^\pm$
3. the phenomenology of the cascade (and other possible damping processes)
4. the value of the background field $B_0$, which enters in the growth rate and also directly affects the level of turbulence $\delta B(k_{\rm res})/B_0$ and the value of $r_L$ that determine $D$ (see Eq. 8).

The typical temporal evolution of the self-generated diffusion coefficient at different distances from the source is shown in Fig. 6 (taken from Ref. [120]) for 10 TeV leptons. At early times, the diffusion coefficient is the background one, while waves start to grow due to the CR driving. At intermediate times $\lesssim 50$–$100$ kyr the diffusion coefficient can be significantly suppressed over a region of $\approx 10$–$20$ pc. At later times, a combination of the decreasing spin-down luminosity and of the turbulent

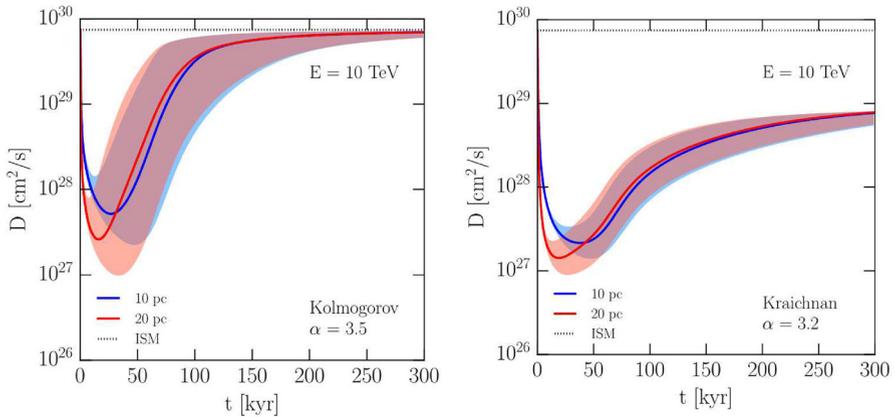

**Fig. 6** Self-generated diffusion coefficient, at distances from the pulsar of 10 pc (blue) and 20 pc (red), as a function of the pulsar age and for particle energy $E = 10$ TeV (taken from Ref. [120]). The efficiency is set to $\approx 50\%$. Left panel: Kolmogorov-type cascade and injection spectrum with slope in energy (momentum) $\alpha = 1.5$ (3.5). Right panel: Kraichnan-type cascade and injection spectrum with slope in energy (momentum) $\alpha = 1.2$ (3.2). Solid lines correspond to $B_0 = 1\,\mu\text{G}$, while the shaded regions correspond to a range $B_0 = 0.5$–$2\,\mu\text{G}$ (the smaller (larger) value of $B_0$ corresponds to the upper (lower) border of the shaded areas). Dotted lines correspond to the Galactic diffusion coefficient

cascade produce a relaxation phase during which the diffusion coefficient returns to the background value. In these plots, it was assumed $R = 1$ pc, a conversion efficiency $\xi_{SD} \approx 50\%$ and hard injection spectra with $\alpha \sim 1.2$–$1.5$ and $E_c \sim 100$ TeV. The effect of the phenomenology of the cascade appears significant, especially in the relaxation phase, with the Kraichnan case relaxing much more slowly than the Kolmogorov case. The value of $B_0$ is also very relevant, as illustrated by the shaded regions which correspond to a range $0.5$–$2\,\mu\text{G}$, with larger $B_0$ corresponding to less pronounced suppression and faster relaxation. This is a typical feature of self-generated diffusion, since, for a given value of $\delta B$, the level of turbulence decreases when $B_0$ increases.

The effect of changing the acceleration efficiency can be seen in the left panel of Fig. 7 (taken from Ref. [121]), where the diffusion coefficient for 10 TeV particles (injection with $\alpha \sim 1.5$ and $E_c \sim 100$ TeV) at a distance of 10 pc from the pulsar is shown for $\xi_{SD}$ ($\xi$ in the figure) ranging between 1 and 100%, and $R = 1$ pc. Clearly, an increase of the efficiency leads to a larger suppression of the diffusion coefficient. Instead, looking at the right panel of the same figure, we can see that the same level of suppression can be obtained with a larger transverse size of the flux tube, provided that the $\propto 1/R^2$ decrease of the injection term is compensated by a corresponding increase in $\xi_{SD}$. When $R$ is assumed to be larger than a few pc, the suppression of $D$ becomes modest and eventually disappears even for an efficiency close to 100%. An even more dramatic impact on the effectiveness of the suppression comes from changing the geometry of particle escape from 1D (flux tube) to spherical: in this case the required level of suppression is impossible to obtain even for particle luminosity equal to 100% of the pulsar $\dot{E}$.

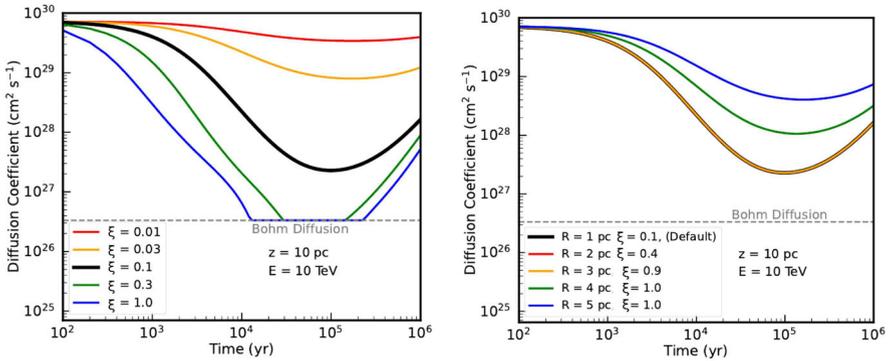

**Fig. 7** Self-generated diffusion coefficient at distances of 10 pc as a function of the pulsar age particle and for energy $E = 10$ TeV, adapted from Ref. [121]. Left panel: hard injection in energy ($\alpha = 1.5$) and varying injection efficiency $\xi = 0.01–1$. Right panel: hard injection in energy ($\alpha = 1.5$) and different combinations of efficiency $\xi$ and flux tube radius $R$

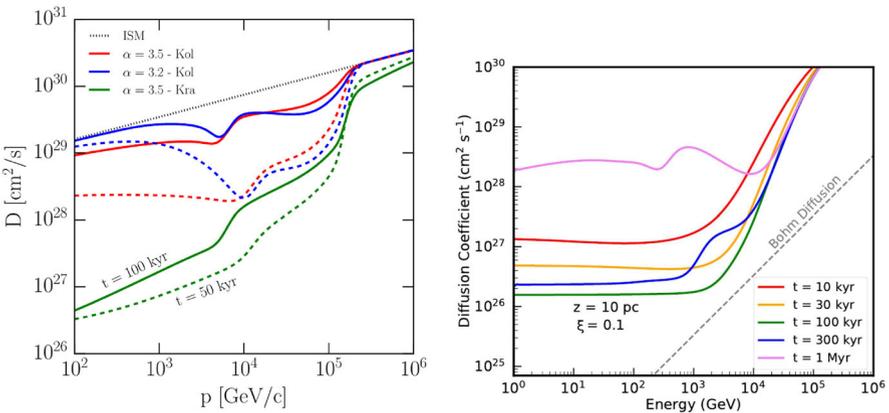

**Fig. 8** Left panel: self-generated diffusion coefficient as a function of the particle momentum at a distance of 10 pc from the pulsar, for ages of 50 kyr (dashed lines) and 100 kyr (solid lines), taken from Ref. [120]. The efficiency is $\approx 50\%$. Red (blue) lines correspond to a Kolmogorov phenomenology and injection spectrum with slope in energy (momentum) $\alpha = 1.5$ (3.5) ($\alpha = 1.2$ (3.2)), while green lines correspond to a Kraichnan phenomenology with $\alpha = 1.5$ (3.5). Right panel: self-generated diffusion coefficient as a function of the particle momentum at distance of 10 pc from the pulsar, adapted from Ref. [121]. The slope of the injection spectrum is fixed to $\alpha = 1.5$ and the efficiency to $\xi = 0.1$. The various lines correspond to different ages of the pulsar

In summary, the diffusion coefficient inferred by the HAWC and LHAASO collaborations is challenging to explain as a result of self-generated turbulence even in the case of 1D transport. In order to obtain the required suppression of $D$ ($D \sim 5 \times 10^{25}$ cm$^2$/s from HAWC data for particles of 10–200 TeV and up to hundreds of kyr, see column 8 in Table 1 and Fig. 8), either a very large efficiency ($\xi_{SD} \sim 100\%$) of conversion of the pulsar spin-down luminosity in released particles or a tiny size ($R \lesssim 1$ pc) of the injection region are required. The latter condition would not be unreasonable in itself, being the size of the nebula exactly in that range, but such a reduced transverse size would unavoidably lead to a non-spherical, elongated morphology of the TeV emission, unless the observer's LOS is very nearly parallel to the local magnetic field direction. In this case, however, it is not clear whether suppressed diffusion is really needed, or the observations can just be explained within the framework of anisotropic transport. In addition, the inclination angle between the magnetic field and the LOS is now even more stringent than in the case of anisotropic transport with small $M_A$ since the transverse size is assumed to remain the same up to 10$s$ pc distance from the source, rather than progressively increase.

## 6 TeV halos and the cosmic ray positron fraction

In the last few years, the leptonic component of the CR flux has been measured with unprecedented precision up to $\gtrsim$ TeV energies. This has added important pieces to the puzzle of the origin of CRs.

Most notably, the positron fraction was found to grow in the energy range $\sim$ 10–200 GeV [7, 123, 124]. Such trend cannot be explained as a result of CR interactions in the ISM in the standard model of CR propagation and is strongly suggestive of additional sources of primary positrons. Since pulsars are prominent factories of $e^\pm$ pairs, they represented a natural candidate and their possible impact on the positron excess has been thoroughly investigated [12, 13, 41, 105, 125]. Although other possibilities have been proposed in the literature, such as secondary production in SNRs [9, 10, 126] or dark matter (see discussion by Ref. [12] and references therein), currently pulsars are generally agreed to be the most likely origin of the positron excess.

In addition, the positron fraction is observed to flatten to a value of $\sim$ 0.15 at $\sim$ 200 GeV, while the positron flux seems to drop above $\approx$ 400–500 GeV [8, 124]. At the same time, the total lepton spectrum was measured up to a few TeV [127–130] and some evidence that it may extend without a cutoff up to $\sim$ 20 TeV was reported on by Ref. [131].

These findings suggest that the sources of multi-TeV leptons, whatever they may be, mostly produce electrons over positrons. Moreover, the short loss time of multi-TeV leptons (see Eq. 14) in the ISM poses the question on the maximum distance of such sources. For instance, a typical $D(10\,\text{TeV}) \approx 10^{29}$–$10^{30}$ cm$^2$/s would translate in (see Eq. 16) $d_{\max} \approx$ 150–500 pc. It should be noted, however, that recent works focused on fitting all available CR data [132], suggest a larger diffusion coefficient that increases this horizon to 1–2 kpc (see Ref. [41] and references therein). The value of $d_{\max}(E)$ affects the number of sources that may contribute to a given spectral range, and for typical scaling of the diffusion coefficient ($\propto E^{0.3-0.7}$) and energy loss rate ($\propto 1/E$, for synchrotron and ICS), it decreases with the particle energy. This was recognized in a variety of works, that addressed the impact of the source stochasticity on the $\gtrsim$ TeV lepton spectrum [41, 133], or in works that proposed a single source as the major contributor to the multi-TeV electron spectrum [134]. Smaller values of the diffusion coefficient would narrow the diffusion-loss horizon and decrease the average number of contributing sources.

In this context, the emerging evidence of a suppressed diffusion region around a few middle-aged pulsars posed new questions. In the assumption that a low-diffusion scenario is the correct interpretation of these haloes, one of the most pressing problems that arises is how they impact the contribution of the Galactic pulsar population to the positron excess. The answer to this question requires to assess the extension of the low-diffusion zones, and the probability that a pulsar is surrounded by such region.

Before discussing the literature on the TeV haloes–positron fraction connection, it is useful to summarize the main findings of a variety of works that investigated the contribution of PWNe to the positron excess in the standard scenario of CR propagation in the Galaxy [12, 13, 120, 135].

In this class of studies, apart from leptons of secondary origin, electrons are assumed to be accelerated by SNRs and released in the ISM with a steep spectrum $\propto E^{-\alpha_{\rm SNR}}$ ($\alpha_{\rm SNR} \approx 2.2$–$2.4$), while electrons and positrons re injected from pulsars with a hard spectrum $\alpha_{\rm PSR} \lesssim 2$ (plus cutoff) or a broken power-law, as discussed in Sect. 5.1.1.

As illustrated in Sect. 2, $e^{\pm}$ pairs can effectively leak out during the bow-shock phase of pulsars, but hardly before then. This implies that escape starts at ages $\gtrsim 50$ kyr, depending on the SNR environment and pulsar speed. This is usually taken into account in the modeling of the positron flux from pulsars, and limits the amount of spin-down energy that can be channeled into high-energy leptons than can reach us. Indeed, given the typical time dependence of the spin-down luminosity, most of the rotational energy is lost when the pulsar is inside the remnant.

Before the halo discovery, it was shown, by several works in the literature, that Galactic middle-aged pulsars can easily account for the rising positron fraction, and the specific potential contribution from known and nearby objects, such as Geminga and Monogem, was also estimated. As a typical example, we report in Fig. 9, the calculation by Ref. [12] of the positron flux and positron fraction from Galactic pulsars and from Geminga. More recent works also included the effect of the source stochasticity on this quantities [41, 135]. For instance, Fig. 10 shows the calculation of the positron fraction by Ref. [41], where the shaded area represents the fluctuations induced by the stochasticity of the sources in space and time. As anticipated above, such effect is more prominent at higher energies, due to the progressive narrowing of the diffusion-loss horizon and the consequent decrease of the number of potentially contributing sources, as shown in the left panel of the same figure.

In this setup, if one assumes that a suppressed diffusion coefficient, similar to that inferred from the HAWC and LHAASO data, is present everywhere in the disk (or in a sufficiently large region in the local space), clearly the lepton flux from known local sources, and in particular from middle-aged pulsars, would be dramatically suppressed. This was shown by the HAWC collaboration itself, that in its seminal paper on Geminga and Monogem [1], estimated the contribution of these two pulsars to the positron flux under this assumption. Since those are the closest middle-aged pulsars, this would strongly question the role of this class of sources in determining the rising positron fraction. On the other hand, a suppression of the diffusion coefficient by a factor 100–1000 would reduce the horizon of $\sim 10$ TeV leptons of a factor 10–30, which means that any source of multi-TeV leptons should be located within the unreasonably small distance $\lesssim 10s$ pc.

For these reasons, several authors suggested that inhibited diffusion may be present only in a region of $\approx 10s$ pc around middle-aged pulsars. Thus, the HAWC and LHAASO data were thoroughly re-examined, together with the *Fermi*-LAT data, in an attempt to re-assess the contribution of this class of sources to the local positron flux in light of the existence of haloes. This class of models, collectively termed as "two-zone diffusion" models, are discussed in more detail below.

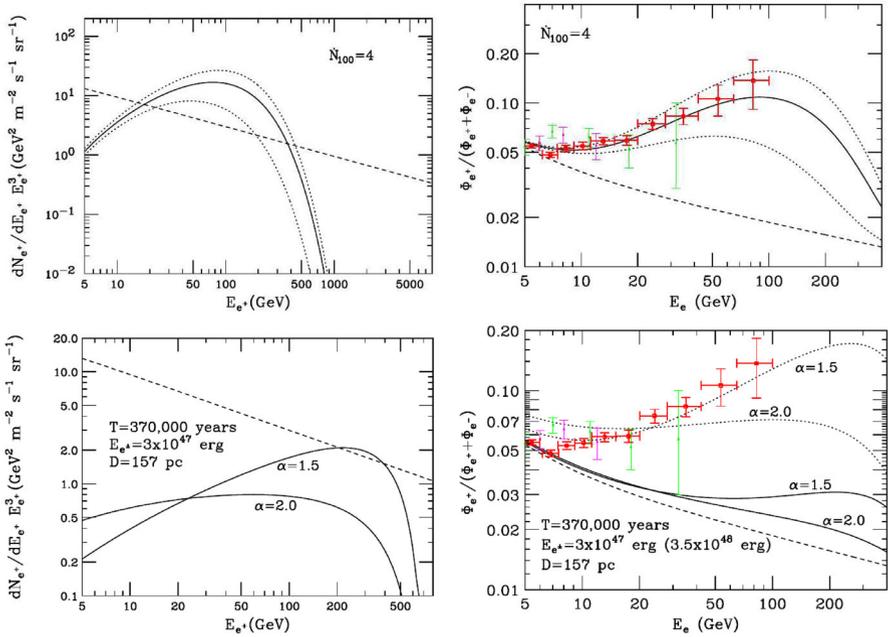

**Fig. 9** Taken from Ref. [12]: spectrum of positrons (left panels) and the positron fraction (right panels, compared to the PAMELA data). The dashed line gives the predicted secondary positron flux. Top panels: contribution resulting from all Galactic pulsars. The injection spectrum is assumed $Q_e \propto \dot{N}_{100}(E/\text{GeV})^{-1.6} \exp(-E/80\,\text{GeV})$, where $\dot{N}_{100}$ is the rate of pulsar formation in units of sources per century. The upper (lower) dotted line represents the case in which the injection rate within 500 pc of Earth is doubled (neglected). Bottom panels: contribution from Geminga. The injection spectrum is assumed $Q_e \propto (E/\text{GeV})^{-\alpha} \exp(-E/600\,\text{GeV})$ with $\alpha = 1.5$ and 2.0. The solid (dashed) lines correspond to a total energy in $e^{\pm}$ of $3.5 \times 10^{47}$ erg ($3 \times 10^{48}$ erg)

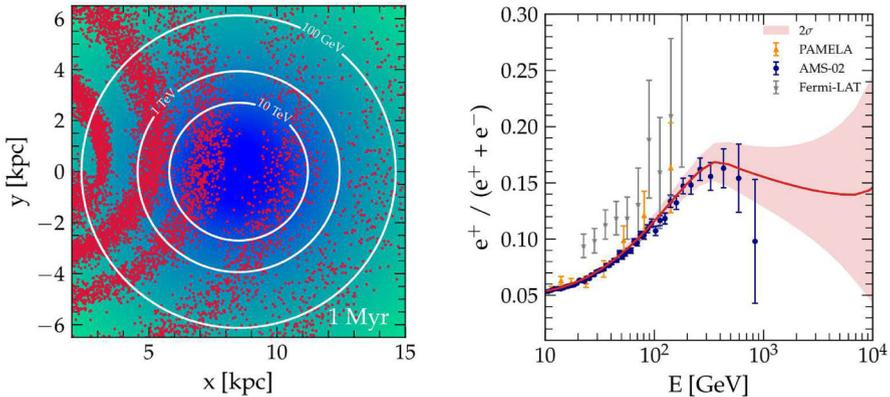

**Fig. 10** Taken from Ref. [41]. Left panel: position of source explosions in the Galactic plane (considering also the spiral structure of the Milky Way) in a given realization and for a simulation time of 1 Myr. The white circles, centered on the location of the Sun, show the horizon for particle energies of 100 GeV, 1 TeV and 10 TeV. Right panel: positron fraction compared with the PAMELA, *Fermi*-LAT, and AMS-02 data. The shaded area shows the effect of fluctuations due to the stochastic nature of sources

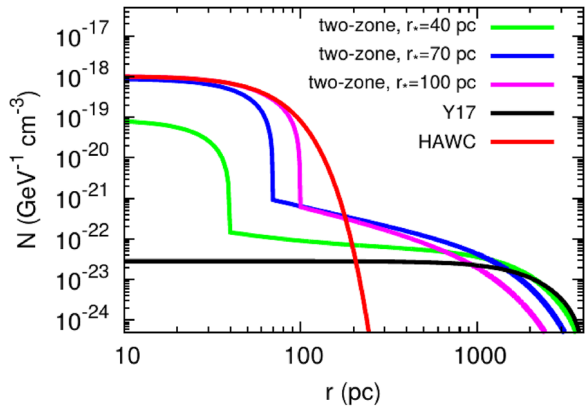

**Fig. 11** Taken from Ref. [139] (© AAS. Reproduced with permission.) Distribution of 1 TeV leptons from Geminga at the present age. The red curve corresponds to a one-zone model with the diffusion coefficient inferred by HAWC [1], while the black curve corresponds to standard Galactic diffusion. The other curves are calculated in the two-zone diffusion model, with different sizes (40 pc, 70 pc, 100 pc) of the low-diffusion zone around the source

### 6.1 Two-zone diffusion models

The two-zone diffusion model typically adopted in the case of pulsar haloes is a modification of the 3D isotropic diffusion setup illustrated in Sect. 5.1, with

$$D(E, r) = \begin{cases} D_{\text{low}}(E) & r \leq R^* \\ D_{\text{ISM}}(E) & r > R^* \end{cases}, \tag{23}$$

namely with a suppressed diffusion coefficient within a radius $R^*$ from the source and the typical ISM diffusion beyond that distance. Variations of this scenario have also been considered with a smoother transition from suppressed to standard ISM diffusion coefficient [136, 137]. An analytic solution for the case of sharp transition was presented in Two-zone diffusion [138] for the case of Geminga, while the other authors solved the transport equation numerically (see, e.g., Ref. [137]).

The particle density as a function of the distance from the source is significantly affected by the presence of a region of suppressed diffusion. This can be well appreciated in Fig. 11 (taken from Ref. [139]), where the particle density at 1 TeV around Geminga is shown in the assumption of burst-like injection for the cases of *i)* one-zone diffusion with suppressed (labeled "HAWC") and ISM (labeled "Y17") diffusion coefficient; *ii)* two-zone diffusion, with transition from "HAWC" to "Y17" diffusion, and $R^* = 40, 70, 100$ pc.

The $e^{\pm}$ distribution resulting from burst injection is

$$f_e^{\text{burst}} \propto r_d^{-3} e^{-\left(\frac{r}{r_d}\right)^2}, \tag{24}$$

where $r_d$ is the diffusion distance of the particles in the time since the burst. For reference, the loss time for 1 TeV leptons is $\tau_{\rm loss} \approx 3 \times 10^5$ yr, which is close to the age of Geminga. The corresponding diffusion radius with the coefficient inferred from HAWC data is $r_d \approx 70$ pc.

Not surprisingly, the case of one-zone suppressed diffusion corresponds to a large density of particles within a few tens of pc around the source. The density, however, drops sharply at $r \sim r_d(D_{\rm low})$. For typical ISM diffusion, the particle density is diluted by orders of magnitudes close to the source (due to the $r_d^{-3}(D_{\rm ISM})$ term), but particles are spread over a much more extended region, so that the corresponding flux beyond $\sim 200$ pc is well above that of the slow diffusion case.

In the two-zone diffusion scenario, particles spend some time in the low-diffusion region, where a large density can be accumulated close to the source. The case $R^* \gg r_d$ is basically indistinguishable from the one-zone suppressed diffusion scenario. When $R^* \gtrsim r_d$ (here for the cases $R^* = 70, 100$ pc) the particle density close to the source (within few tens of pc in this case) is very similar to that of the one-zone scenario with $D_{\rm low}$. From the point of view of interpreting the HAWC data, this would make indistinguishable any scenario with sufficiently large $R^*$. Instead, with a progressive reduction of $R^*$ below $r_d$, the density is reduced even close to the source and eventually approaches that corresponding to standard ISM diffusion when $R^* \ll r_d$.

Eventually, particles enter the fast propagation region and can diffuse much further than in the low-diffusion one-zone case. The density of particles at a given location is the result of two competing effects. On the one hand, a larger value of $R^*$ implies a larger particle density in a more extended region, and typically a larger particle density even beyond $R^*$ (as a result of dilution setting in on larger scales). On the other hand, the effective diffusion radius (taking into account the time spent in the two zones) tends to be smaller with larger $R^*$, which can result in a cutoff closer to the source. A feature that is very important in establishing the contribution of positrons from nearby middle-aged pulsars is that the particle density at distances $\gtrsim 200$–300 pc can be enhanced in the two-zone model compared to the case of typical ISM diffusion.

Obviously, these qualitative results strongly depend on the particle energy, and in particular on the ratio between $r_d$ (which depends on the diffusion coefficient and on the loss time or age) and $R^*$. If $r_d \ll R^*$, it might be impossible for particles to leak out of the low-diffusion region, while if $r_d \gg R^*$ the impact of the low-diffusion region on the CR density would be much reduced.

For instance in Fig. 12, taken from Ref. [137], we report their prediction for the $\gamma$-ray SB around Geminga in the range $E_\gamma = 8$–40 TeV (compared to the HAWC data, corresponding to $E_e \approx 20$–200 TeV) and in the range $E_\gamma = 10$–30 GeV (corresponding to $E_e \approx 1$–10 TeV [140]). The various curves have been obtained with the assumption of stationary injection from Geminga, and with a smooth transition from a low-diffusion region ($r < R^*$) to the normal ISM (beyond $r > R_t$), labeled as $SDZ(R^*, R_t)$.

In order to fit the HAWC data, a minimal condition on $R^*$ is that it should be comparable to or larger than the projected distance covered by the HAWC data, which, for Geminga, corresponds to $R^* \gtrsim 30$ pc. For $E \approx 10$–200 TeV, the value of $r_d$ in the low-diffusion coefficient found by HAWC is $r_d \approx 25$–30 pc. With an appropriate choice of the ratio $D_{\rm low}/D_{\rm ISM}$, the data can be reproduced, and since $R^* \gtrsim r_d$, any further increase of $R^*$ (up to the extreme case of low diffusion everywhere in the disk) will not affect the fit to the HAWC data, but will reduce the positron flux that can reach Earth at $E \gtrsim 10$ TeV.

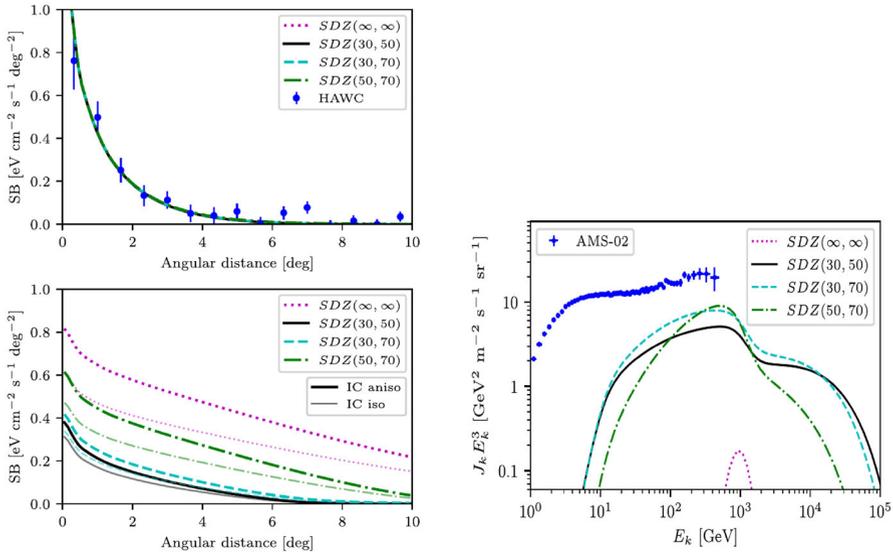

**Fig. 12** Taken from Ref. [137] (© AAS. Reproduced with permission.) Left panel: predicted SB of Geminga. The top panel refers to the energy range $E_\gamma = 8$–40 TeV and the results are compared to the HAWC data; the bottom panel refers to the range $E_\gamma = 10$–30 GeV. The different curves correspond to different sizes of the low-diffusion zone ($R^*$) and of the transition zone ($R_t$) between low diffusion and ISM diffusion, as encoded in the legends $SDZ(R^*, R_t)$. Right panel: contribution of Geminga to the local positron flux in different two-zone diffusion setups (indicated by $SDZ(R^*, R_t)$)

Instead, for $E \approx 1$–10 TeV particles $r_d \approx 35$–70 pc and the SB is strongly affected by the value of $R^*$. For the one-zone case, the SB in this energy range is clearly maximized, while the positron flux is strongly suppressed. The progressive reduction of $R^*$ below $r_d$ reduces the SB, but allows for 1–10 TeV leptons to reach Earth. Higher fluxes are obtained with combinations of $R^*$ and $R_t$ that are not so large as to prevent particles to reach a given location at a given time, and not to small as to let particles escape too fast. Instead, at intermediate values, the accumulation of particles in the low diffusion region can reflect in an enhanced ICS flux (depending also on distance and time).

Based on these considerations, it is clear that, if both LHAASO/HAWC data and *Fermi*-LAT data are available for a given pulsar halo, it may be possible to constrain the injection spectrum, the size of the low diffusion region and the corresponding positron flux. This was thoroughly investigated by Ref. [141], who performed a detailed study using the HAWC and *Fermi*-LAT data by Ref. [67] for Geminga and Monogem (upper limits for the latter). Taking into account the time when pulsars are likely to have left the parent remnant, they assume that a fraction $\xi_{SD}$ of the pulsar spin-down luminosity is converted in $e^\pm$ with a broken power-law spectrum, with slope $\alpha_1 = 1.5$ below the break $E_b = 100$ GeV and $\alpha_2 = 2.2$–$2.8$ above the break. $R^*$ is varied in the range 30–120 pc.

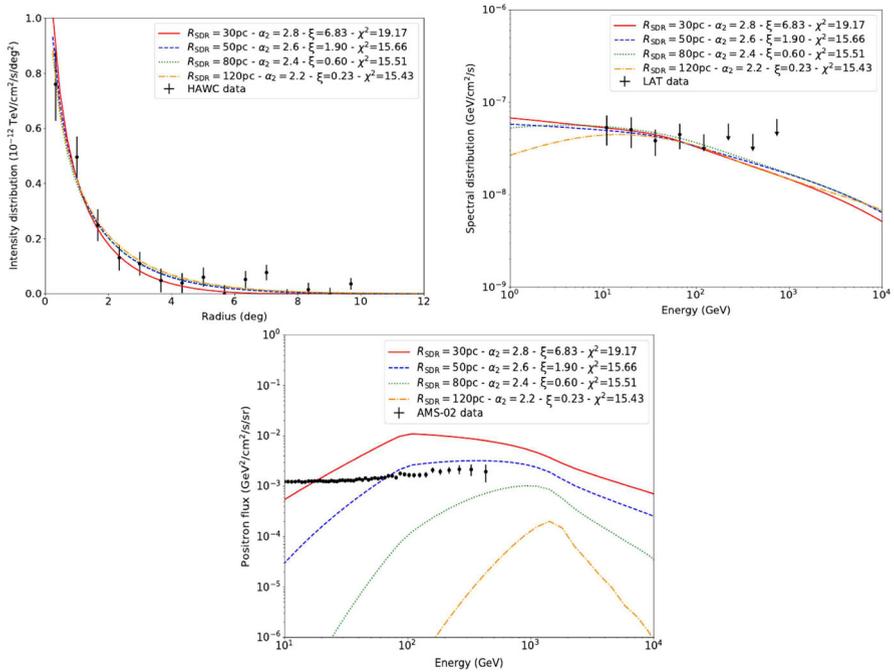

**Fig. 13** Adapted from Ref. [141]. Predicted SB of Geminga in the HAWC (top left panel) and *Fermi*-LAT (top right panel) range. The injection spectrum is assumed as a broken power-law with $\alpha_1 = 1.5$ below the break ($E_b = 100$ GeV) and varying $\alpha_2 = 2.2$–$2.8$ above the break. $\xi$ marks the conversion efficiency needed for a given injection spectrum to reproduce both the HAWC and *Fermi*-LAT data. Bottom panel: the positron flux at Earth corresponding to the different choices of parameters is shown

Under these assumptions (and a few other discussed in the paper), in the case of Geminga, the large *Fermi*-LAT flux reported by [67] implies that smaller values of $R^*$ require larger values of $\alpha_2$ (steeper spectrum above the break) in order to fit the data. Such combinations also need $\xi_{SD} \gg 1$ and result in a too large flux of positrons at Earth. Without changing any other parameter, in order to get reasonable values of $\xi_{SD}$ and not overshoot the positron flux, the requirements is that $R^* \gtrsim 80$ pc for Geminga (see Fig. 13).

The case of Monogem is less constrained, since only upper limits in the *Fermi*-LAT data were reported. In this case, $R^* > 30$ pc can provide both a reasonable fit to the HAWC data, without overshooting the low-energy upper limits, and to the positron flux.

Interestingly, in this setup, both Geminga and Monogem require an acceleration efficiency $\gtrsim 30$–$40\%$ even for large $R^*$, in order to simultaneously match both HAWC and *Fermi*-LAT data. This result was used by Ref. [141] also to figure out how rare low-diffusion haloes around middle-aged pulsars should be. Indeed, they estimated that, as shown in Fig. 14, if all known middle-aged pulsars within 1 kpc develop a halo

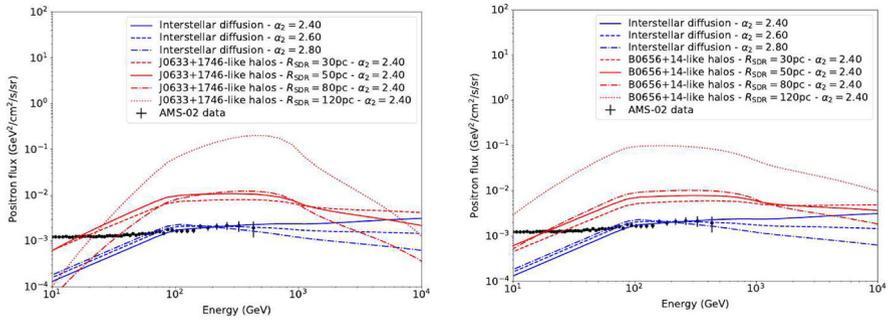

**Fig. 14** Taken from Ref. [141]. Total local positron flux from known middle-aged pulsars within 1 kpc computed in the case in which they develop a halo like Geminga (left panel) or Monogem (right panel) after an age of 60 kyr. Geminga, Monogem and B0656+14luded from the calculation. Different sizes ($R_{\text{SDR}} = R^*$) of the low-diffusion zone are considered. For comparison, the flux in the case of typical ISM diffusion is also shown for varying spectral slope above the break, $\alpha_2$. In all cases, an injection efficiency of 100% was assumed

similar to that of Geminga and Monogem, they would overshoot the observed positron flux unless the acceleration efficiency is < 10–30%. Instead, if none of these pulsars develop a low-diffusion halo they would saturate the positron flux with an average injection efficiency of 70–100%. They conclude that the occurrence of low-diffusion haloes could be as low as 5–10%, although these numbers may change a bit, taking into account the full parameter space (see Ref. [141] for a detailed discussion).

## 7 Summary and future prospects

In this article, we have tried to provide an overview of the current status of our understanding of the phenomenon of TeV haloes around pulsars. The amount of studies that have appeared in the literature around this subject is striking, given that for years this class of sources only encompassed two objects, and only recently, a third one has been added. The reason for the scientific excitement around this phenomenon is that, as we have extensively discussed, it has far-reaching implications, ranging from CR transport in the Galaxy to the determination of the pulsar contribution to CR positrons at Earth.

The latter issue is especially critical and of interest for a much broader community than that of High Energy Astrophysics, since it is tightly related to the question of whether there is need, or room, for an additional, possibly non-astrophysical, but rather dark matter related, contribution to explain the positron excess. While the halo phenomenon is still not fully understood, current studies seem to converge on a broadly agreed conclusion at least on this aspect: the halo phenomenon does not dramatically affect the pulsar contribution to Earth detected positrons.

This does not make haloes uninteresting from the CR physics point of view. They still provide an excellent means to investigate CR transport in the disk directly, probing the particle distribution in the cleanest possible way, namely through their interaction with the universal and uniform CMB. The situation is very different than what is typically the case for hadronic CRs or for lower energy leptons, that can only be observed when interacting with non-uniform and poorly known distributions of matter, magnetic field and other radiation fields. In a sense, multi-TeV leptonic emission is the only diagnostics in astrophysics that takes us close to the particle detectors employed in space physics, and it seems mandatory to exploit it at best.

Studies focused on understanding particle transport in haloes have so far proven inconclusive in terms of assessing whether anisotropic or isotropic but suppressed transport is at work, nor is it clear, in the latter case, what the origin of the needed suppression might be, whether intrinsic, due to the instabilities induced by the particles escaping their sources, or environmental, due to the peculiar location of the detected sources. Part of the problem arises from the paucity of objects so far detected, and the solution of the puzzle is likely to require an increase of the class members.

After the initial detection of Geminga and Monogem by HAWC [1], a few different research groups [122, 141–143] engaged in the effort of predicting the abundance of this new class of sources in the Galaxy and the prospects for their detection at VHE or in other wavebands. The results of these different studies are *very diverse, predicting* a number of haloes ranging from very few [141] (essentially consistent with *the expectation of no future detections*) to hundreds [142]. In fact, any prediction on the abundance of pulsar haloes is model dependent, and currently very difficult to make in a reliable way, for lack of a broadly agreed interpretation of the phenomenon.

Even more difficult to estimate is the number of haloes that current and upcoming VHE experiments will observe. Haloes can have in principle rather large sizes and the detection of extended sources is a most challenging task at Very High Energies. LHAASO is currently exploring the UHE sky with unprecedented sensitivity and constantly finding new sources as its photon statistics increases. Its observations have already revealed an unexpectedly high level of diffuse emission in the galactic plane [6], that could in principle be due to the superposition of many unresolved haloes. With accumulating observation time, LHAASO might be able to firmly identify more haloes if they are indeed common.

In the TeV range, the sensitivity of LHAASO worsens and an actual leap in the potential of discovery of new sources will only come with CTA [144]. Currently, the CTA capability in terms of halo detection is difficult to assess [145], being strongly dependent on their sizes, which in turn depend, in current models, on the degree of suppression of the diffusion coefficient. In the wait for CTA, ASTRI Mini-Array [146] will serve as a pathfinder: with its large field of view and its ability to collect photons with energies up to 300 TeV, this experiment is expected to bring important new insight on the spatial and spectral properties of the halo surrounding Geminga [147]. Observations of Geminga and Monogem with ASTRI Mini-Array will also serve as a test bench to improve the IACT analysis tools for extended emission in view of CTA, and allow a more realistic prediction of the constraints that CTA observations will be able to set, either through detection or non-detection of additional haloes.

In extreme summary, while the initial driver for excitement around Pulsar haloes, namely the prospect of excluding a pulsar related origin of the positron excess and opening the way for dark matter interpretation, has now waned, these sources offer an unprecedented opportunity to study the propagation of accelerated particles immediately after injection in the ISM. In order to improve our understanding, more observations are needed together with more detailed modeling of particle propagation in the near source environment and possible non-linear effects. To characterize the latter, multi-wavelength data will be an essential addition to the improved capabilities of existing and upcoming Very and Ultra High energy experiments.

**Acknowledgements** The authors acknowledge P. Blasi, C. Evoli, S. Gabici, G. Morlino and B. Olmi for the many useful discussions on the subject of this review.

**Author Contributions** Both the authors have contributed to the writing and approved the manuscript in its final version.

**Funding** Open access funding provided by Istituto Nazionale di Astrofisica within the CRUI-CARE Agreement. This work has been funded by the European Union—Next Generation EU, RFF M4C2 1.1 through grant PRIN-MUR 2022TJW4EJ. EA also acknowledges support from INAF through grant PRIN-INAF 2019.

**Data availability statement** Data availability is not applicable to this article as no new data were created or analyzed in this study.

## Declarations

**Conflict of interest** The authors declare no conflict of interest.